\documentclass[amssymb]{revtex4-2}
\usepackage[utf8]{inputenc}
\usepackage{url}
\usepackage{graphicx}% Include figure files
\usepackage{bbm}
\usepackage{colortbl}
\usepackage{xcolor}
\usepackage{amsmath}
\usepackage{hyperref}
\usepackage{chngcntr}
\usepackage{multirow} 
\usepackage{soul}

\usepackage[capitalise]{cleveref}

\begin{document}

\newcommand{\beginsupplement}{%
        \setcounter{table}{0}
        \renewcommand{\thetable}{S\arabic{table}}%
        \setcounter{figure}{0}
        \renewcommand{\thefigure}{S\arabic{figure}}%
        \setcounter{section}{0}
        \renewcommand{\thesection}{S\arabic{section}}%
     }
\newcommand{\as}[1]{\textcolor{teal!80!green} {{#1}}}

\newcommand{\ik}[1]{\textcolor{purple} {{#1}}}

\title{Combined topological and spatial constraints are required to capture the structure of neural connectomes}
\author{Anastasiya Salova}
\affiliation{Department of Physics and Astronomy, Northwestern University, Evanston, IL 60208}

\author{Istv\'an A. Kov\'acs}
\affiliation{Department of Physics and Astronomy, Northwestern University, Evanston, IL 60208}
\affiliation{Northwestern Institute on Complex Systems, Northwestern University, Evanston, IL 60208}
\affiliation{Department of Engineering Sciences and Applied Mathematics, Northwestern University, Evanston, IL 60208}
\email{istvan.kovacs@northwestern.edu}

\date{\today}

\begin{abstract}

Volumetric brain reconstructions provide an unprecedented opportunity to gain insights into the complex connectivity patterns of neurons in an increasing number of organisms. Here, we model and quantify the complexity of the resulting neural connectomes in the fruit fly, mouse, and human and unveil a simple set of shared organizing principles across these organisms.
To put the connectomes in a physical context, we also construct contactomes, the network of neurons in physical contact in each organism. With these, we establish that physical constraints---either given by pairwise distances or the contactome---play a crucial role in shaping the network structure. For example, neuron positions are highly optimal in terms of distance from their neighbors. However, spatial constraints alone cannot capture the network topology, including the broad degree distribution. Conversely, the degree sequence alone is insufficient to recover the spatial structure. We resolve this apparent conflict 
by formulating scalable maximum entropy models, incorporating both types of constraints. The resulting generative models have predictive power beyond the input data, as they capture several additional biological and network characteristics, like synaptic weights and graphlet statistics.

\end{abstract}

\maketitle

\section{Introduction}

%Representing the brain as a complex network 
Network representations of the brain offer a key to relating
its multiscale structure to its dynamics and function \cite{telesford2011brain,barabasi2023neuroscience,bassett2017network}. 
The widespread availability of the data at the scale of inter-regional and inter-areal connections \cite{scannell1995analysis,kotter2004online,lanciego2011half,markov2014weighted,oh2014mesoscale,zingg2014neural,chiang2011three}
made it possible to distill fundamental design principles of brain organization
\cite{bullmore2009complex,betzel2019distance,horvat2016spatial,bullmore2012economy}.
 Such macroscopic and mesoscopic networks must emerge from cellular-level synaptic networks (neural \textit{connectomes}) where nodes represent individual neurons and edges correspond to the presence of synaptic connections between pairs of nodes.  
Yet, up until recently, analyses of neural connectomes have been severely limited in their scope
\cite{bassett2010efficient,chen2006wiring,maertens2021multilayer,yan2017network,gushchin2015total,song2021maximum}. 
The advent of volumetric brain reconstructions opens the possibility to 
map and model brain networks in a bottom-up manner \cite{lynn2024heavy,haber2023structure}.
Recent advances and major collaborative efforts in experimental, image analysis, and machine learning techniques have led to an unprecedented amount of nanometer-resolution 
brain datasets that span a variety of organisms  \cite{helmstaedter2013cellular, hildebrand2017whole, motta2019dense, winding2023connectome, dorkenwald2023neuronal}.

\begin{figure}
    \centering
    \includegraphics[width=1.0\linewidth]{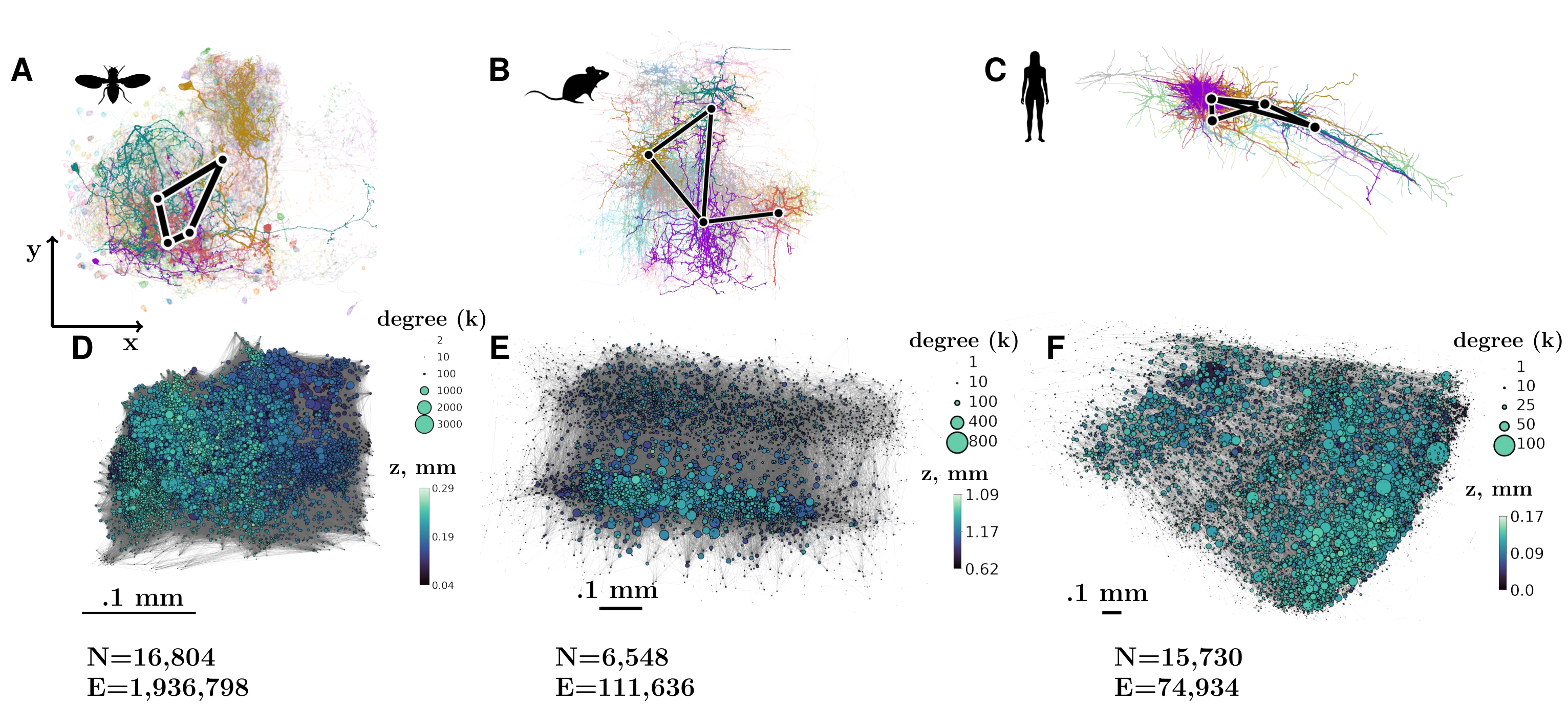}
    \caption{Visualization of the connectomes in the $xy$ plane (axes shown on the left). \textbf{A-C}: 
    sub-networks of fruit fly, mouse, and human neural connectomes. Neurons corresponding to network nodes (black dots) are shown in bright colors. A subset of neighbors of these neurons are shown in dimmer colors. \textbf{D-F}: full connectomes. Node positions correspond to the location of neuron centers of mesh, and node color and size correspond to their undirected degree $k$ and $z$ position, respectively. Edges are shown in light grey. We also provide the number of nodes ($N$) and edges ($E$) in each network.}
    \label{fig: basic properties}
\end{figure}

Here, we aim to quantify and compare the neural connectomes across organisms and understand the common design principles that lead to their complex structure.
We focus on the roughly millimeter-scale connectomes of the adult fly, mouse, and human brains %limit our analysis to connectomes of three of the 
 %containing  %neurons and non-neuronal cells
\cite{xu2020connectome,bae2021functional,shapson2021connectomic}. These datasets allow us to compare the microscopic structure of the human brain to mammalian and even non-mammalian brains \cite{van2016comparative}. 
As an invertebrate model, we use the fruit fly hemibrain dataset, capturing most of the central brain of the adult female \textit{Drosophila melanogaster} \cite{xu2020connectome}.
As a mammalian model, we use the male mouse brain dataset that spans multiple cortical visual areas 
\cite{bae2021functional}. 
Finally, we use the 
reconstruction of a sample from the temporal lobe of the cerebral cortex in the female human brain \cite{shapson2021connectomic} to analyze the microscopic structure of the human connectome. 
We consider the undirected unweighted versions of the connectome obtained from these three datasets, where each edge represents the presence of some chemical synapses. The basic properties of these networks, each containing thousands of nodes and up to millions of edges (see \cref{fig: basic properties}), are summarized in \cref{table: basic info}. 
The fruit fly neural connectome comprises $13\%$ of the approximately 120 thousand fruit fly neurons \cite{dorkenwald2023neuronal}. In comparison, the mouse and human connectomes contain much smaller fractions of roughly 100 billion neurons in the human brain \cite{azevedo2009equal} and 100 million neurons in the mouse brain \cite{herculano2006cellular}.

To start, connectomes are inherently spatial \cite{barthelemy2011spatial,bassett2018spatial,markov2011weight,horvat2016spatial,markov2013role}. 
Neurons %including the ones in the connectomes we analyze, 
are indeed complex fractal-like objects \cite{smith2021neurons,ansell2023unveiling} embedded in three-dimensional space. These neurons form complex \textit{physical} networks---networks where nodes (neurons) of complex shapes can not overlap \cite{pete2023network,posfai2023impact}. Connectomes are also expected to be shaped by a balance of minimizing the wiring cost while maintaining complex topology, at least at large enough scales \cite{bullmore2012economy}.
Yet, there are several key open questions to address at the level of the neural connectome, such as: i) Are the neural connectomes complex networks in the usual sense, showcasing a scale-free degree sequence and the small-world property? ii) To what extent is the position of a neuron given by its neighbors, i.e., are the spatial patterns determined by the topology? iii) Are there signs of optimal wiring given the neuron positions, i.e., to what extent is the connectome dictated by the spatial aspects? iv) Can we design generative network models that capture the main topological and spatial features of neural connectomes?

\begin{figure}[h]
    \centering
    \includegraphics[width=.7\linewidth]{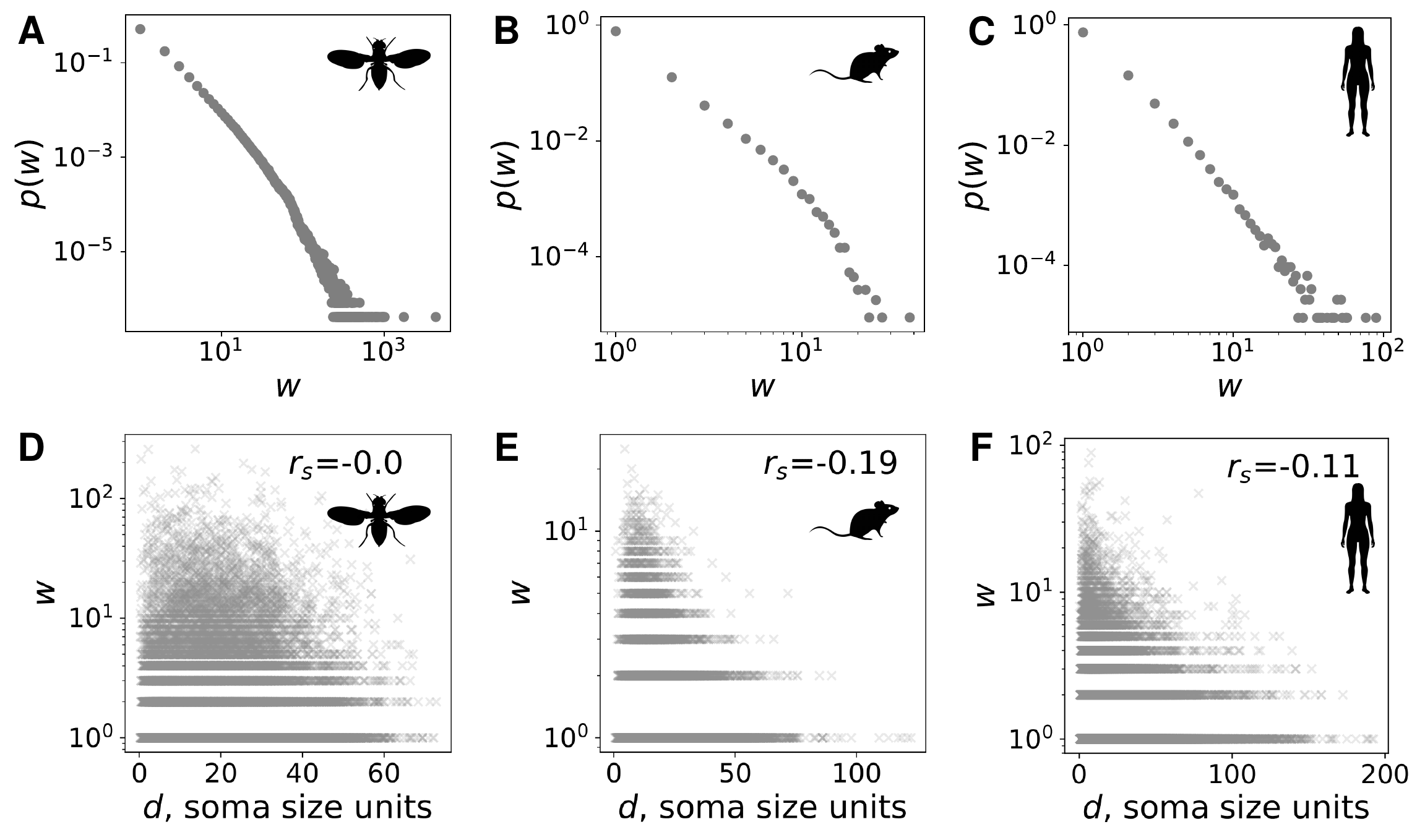}
    \caption{Degree distribution and distance dependence for edge weights. Here, edge weight is defined as the number of synapses between a given pair of neurons. Synapse strength can also be quantified by its size in mammalian brains, thus other definitions of edge weight could also be appropriate. \textbf{A-C}: distribution of edge weights. \textbf{D-F}: edge weight as a function of distance.}
    \label{fig: weighted distributions}
\end{figure}

Broad degree distributions are essential ingredients of complex networks, including biological networks \cite{amaral2000classes,barabasi1999emergence,lynn2024heavy,lynn2022emergent}. As a first observation about the neural connectome topology, we note that the degree distributions of these networks, shown in \cref{fig: randomizations} D-F, are broad but not
scale-free. This contrasts with the distributions of edge weights---here, defined by the number of synapses between each pair of neurons--- that are heavy-tailed consistently with the findings of Ref.~\cite{lynn2024heavy}, as shown in \cref{fig: weighted distributions} A-C.
We also show that given the topology, the total wiring length---as quantified by the sum of Euclidean distances between all the pairs of connected neurons---is highly optimal compared to %spatial connectome randomizations obtained by 
versions with randomly shuffled node positions.

To gain further insights, we \emph{hypothesize} that the biological mechanisms required to form and maintain synapses might be different from those allowing neurons to be in physical contact. While synapses can be only observed between neurons in physical contact, physical contact alone might not dictate synapse formation, a parts of neurons could get in close proximity in passing. 
%If this is true, then it can be a useful tool to 
To test the impact of physical contact on the connectome, we construct the neural \emph{contactome} of each dataset (for details, see Methods \ref{subsec: contact}), summarizing which neuron pairs are within a distance allowing synapse formation \cite{peters1976projection,braitenberg1991peters,rees2017weighing,stepanyants2005neurogeometry, reimann2015algorithm, kovacs2020uncovering}.
By construction, neural connectomes are sub-networks of the neural contactomes, as illustrated in \cref{fig: multiplex network} A using a subset of human neurons. 

As a spatial observation, we demonstrate that the edge probability in our millimeter scale connectomes decays exponentially with distance, as shown in \cref{fig: randomizations} A-C. This is in line with the findings of exponential distance dependence in the \textit{C. elegans} connectome \cite{arnatkevicute2018hub}, mouse neural cell cultures \cite{yin2020network}, and inter-areal level connectomes in mammalian brains \cite{fulcher2016transcriptional,fornito2019bridging}. 
We find a similar exponential distance dependence in the contactome (see \cref{fig: multiplex network} B and \cref{table: orientation}).
However, connectomes are not a random subset of contactomes, as indicated by the nonlinear relationship between the connectome and contactome degrees of individual neurons (see \cref{fig: multiplex network} bottom row insets). Thus, the contactome constraints alone are insufficient to obtain accurate connectome models, confirming our hypothesis. 
Moreover, the exponential distance dependence alone is insufficient to capture the topology of the neural connectome, even at the level of recovering the broad degree distribution. Similarly, the degree sequence does not imply the correct form of distance dependence. Thus, there is an apparent conflict between network topology and distance dependence.

Our analyses suggest that a synergistic combination of spatial (e.g., contactome or distance dependence) and topological (e.g., degree sequence) constraints is required to form realistic models of neural connectomes.
To incorporate these two types of constraints, we develop a range of scalable generative models using canonical maximum entropy ensembles \cite{park2004statistical,bianconi2021information,robins2007introduction,dichio2023statistical}.   
The combination of the intrinsically probabilistic nature of maximum entropy models and their ability to preserve local and global constraints makes them versatile for representing neural connectomes.
Conceptually, such a  framework is capable of capturing stereotypic brain connections together with individual variability %and technical noise present 
in connectome datasets \cite{witvliet2021connectomes,hiesinger2018evolution,schlegel2023whole}. In addition to preserving average quantities---soft constraints--- we utilize the ability of maximum entropy models to respect hard constraints \cite{kovacs2020uncovering,hao2023proper} by considering a class of models that only allow the formation of synapses between neurons in physical contact, see \cref{subsec: contact constraint}. 

Models that aim to accurately represent the connectome need to capture its structure beyond the built-in network constraints. We show that the maximum entropy models that preserve distance and degree-based network features match other network properties, such as clustering, graphlet counts, and measures related to the length of shortest paths, see \cref{fig: fold changes,fig: fold changes extra}. 
Interestingly, our maximum entropy models %edge probabilities obtained from the maximum entropy models appear to 
have predictive power beyond what is dictated by the input data, as illustrated by the correlation between the edge probabilities in the models and the synaptic weights %, as indicated by their correlation 
(see \cref{fig: weight correlation}). Additionally, models that include contact constraints capture the heterogeneity in distance dependence associated with neuron alignment in the mouse and human cortex, as shown in \cref{fig: neuron orientation fold change}. Altogether, we demonstrate the ability of simple network-based models to faithfully represent the connectomes across species, even in the absence of detailed organism-specific biological information.

\section{Results}

\subsection{Connectome models incorporating distance dependence and degree sequence} \label{subsec: null}

\begin{figure}
	\centering
	\includegraphics[width=.9\linewidth]{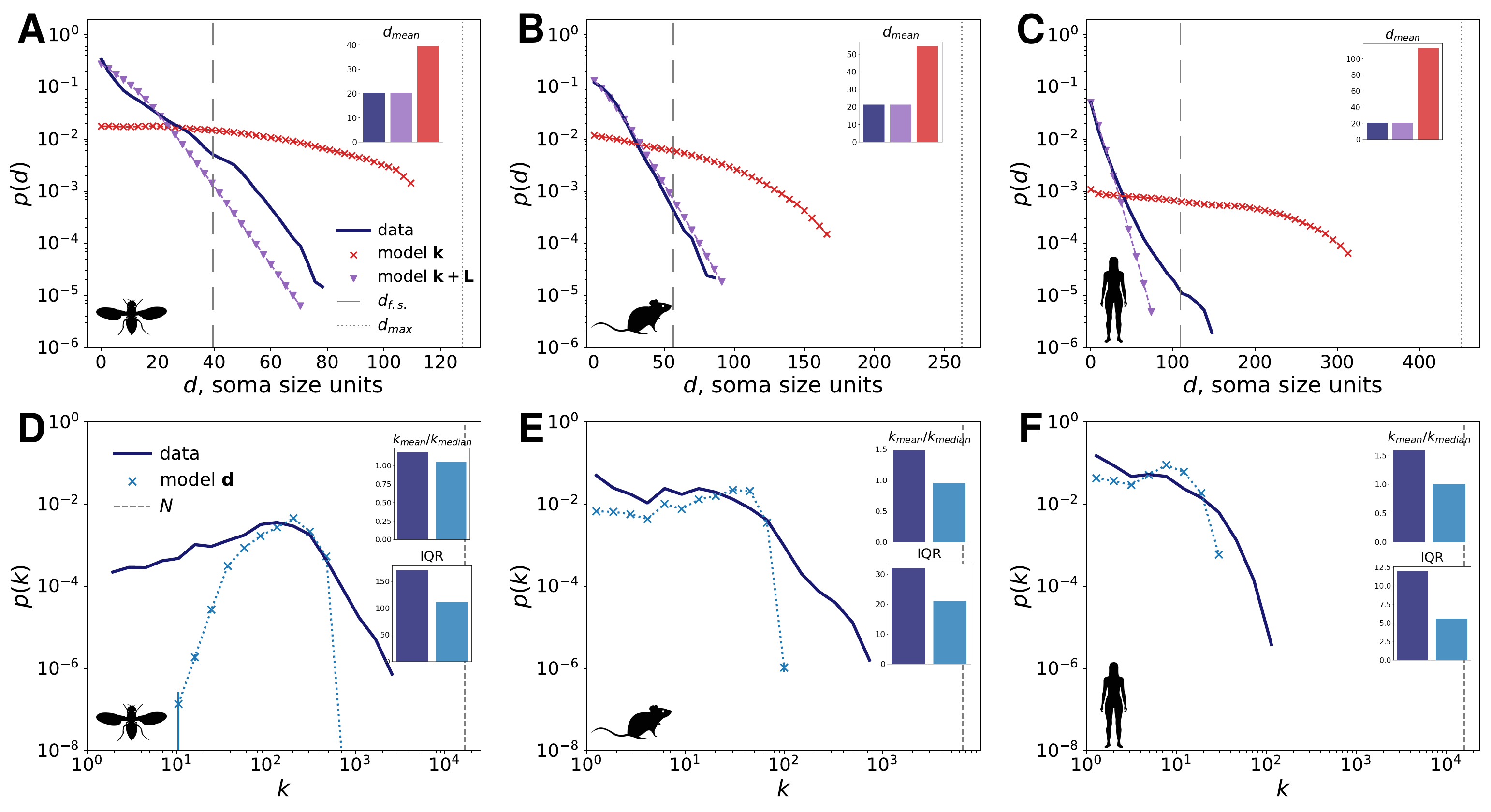}
	\caption{Distance and degree distribution of the connectome data and models. For a definition of soma size units used in the $d$ axis, see Methods \ref{subsec: soma sizes}. Each model distribution shows the average between 100 realizations, and the error bars represent standard deviations. \textbf{A-C}: distance distribution of the synaptic network data (dark blue), model \textbf{k} (red) and model \textbf{k+L} (purple). The vertical line at $d_{\text{max}}$ corresponds to the largest distance between the neurons in the dataset. The vertical line at $d_{\text{f.s.}}$ corresponds to the distance between neurons above which the finite size of the experimental volume affects the distance dependence. This distance corresponds to the peaks of the pairwise neuron distance distributions shown in grey in \cref{fig: distance densities}. \textbf{A-C insets}: mean inter-neuron distance between the connected pairs in data (dark blue) and models.
		\textbf{D-F}: 
		degree distribution of the synaptic network data (dark blue) and model \textbf{d} (light blue).
		\textbf{D-F, upper insets}: ratio of mean degree to median degree in data and model \textbf{d}.
		\textbf{D-F, lower insets}: interquartile range (IQR) of the degree distribution of data and model \textbf{d}.
	}
	\label{fig: randomizations}
\end{figure}

To characterize the spatial organization of connectomes, we model each neuron as a node whose position corresponds to the ``center'' of that neuron, given by the ``centers of mesh'' of its surface, see Methods \ref{subsec: spatial aspects}. We then use the Euclidean distance between the neuron positions to represent the relative locations of pairs of neurons. As our first result, we find that the probability of forming an edge of the connectome decays rapidly as a function of distance. 
Qualitatively, the decay appears exponential, as demonstrated in the top row of \cref{fig: randomizations}. Quantitatively, assuming exponential decay of a form $p(d)\propto e^{-d/d_0}$, 
we estimate the constant $d_0$ for the mouse connectome to be $\sim$0.05 mm. This scale differs from the scale of the exponential decay in inter-areal connectivity by an order of magnitude---$d_0$ is estimated to be 2.179 mm for the mouse in Ref.~\cite{fulcher2016transcriptional}, which is larger than the linear dimensions of the mouse dataset studied here.
%SI Fig S2 A
To compare the characteristic distances $d_0$ of the three organisms, we use the soma size as a distance scale. Estimating the soma sizes is discussed in \cref{subsec: soma sizes}, and their numerical values are presented in \cref{table: basic info}. We estimate $d_0$ to be 10, 10, and 12 in terms of some size units for fly, mouse, and human. %These values are of the same order of magnitude, supporting soma sizes as a useful distance unit.
Therefore, soma sizes set a distance unit that brings the characteristic network distances to the same scale.

\begin{figure}
    \centering
    \includegraphics[width=.9\linewidth]{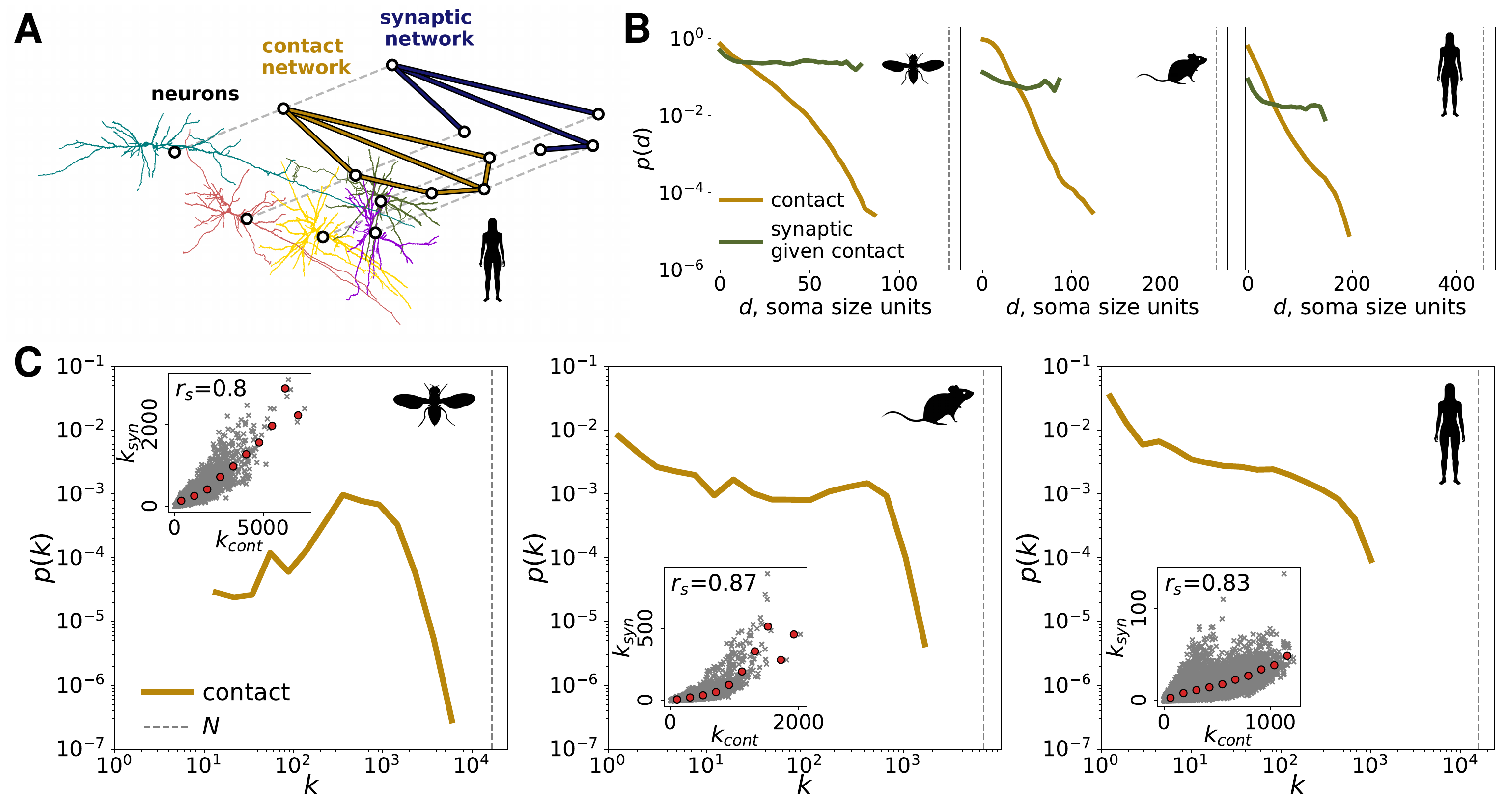}
    \caption{Relationship between the connectome and the contactome. \textbf{A}, from left to right: $xy$ projection of mesh vertices of 5 human neurons and their center of mesh positions (black circles).
    Spatial contactome, or contact network, edges shown in dark yellow. Spatial connectome, or synaptic network, edges are shown in dark blue. The synaptic network and contact network have the same nodes, while synaptic
    edges are a subset of the contact network edges. The resulting network is naturally multilayer (multiplex) \cite{kivela2014multilayer}, as illustrated by connecting the representations of the same nodes with dashed lines.
    \textbf{B}: distance distribution of the contactome (dark yellow) and distance distribution of the connectome restricted to contactome edges (dark olive). Only the data points corresponding to at least 10 edges at a given distance bin are shown.
    \textbf{C}: degree distribution of the contactome. \textbf{C, insets:} degree of the node in the contactome (\textit{x}-axis, $k_{\text{cont}}$) vs its degree in the connectome (\textit{y}-axis, $k_{\text{syn}}$), shown in grey. Red circles represent the average values of $k_{\text{syn}}$ corresponding to a binned range of $k_{\text{cont}}$. We also provide the Spearman's rank correlation coefficient $r_s$ between the $k_{\text{cont}}$ and $k_{\text{syn}}$ variables.}    \label{fig: multiplex network}
\end{figure}

\begin{figure}
    \centering
    \includegraphics[width=.8\linewidth]{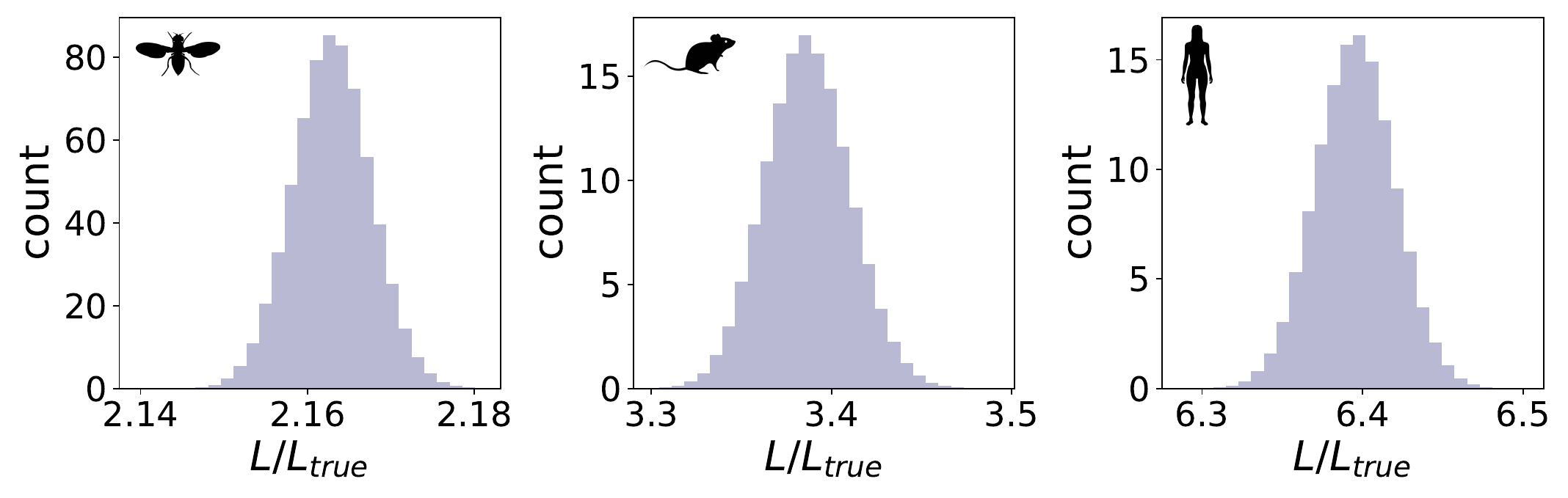} \caption{Distribution of total wiring lengths obtained by randomly shuffling the neuron positions while fully preserving the network topology. 200,000 randomized spatial networks were obtained for each organism.}
    \label{fig: shuffle}
\end{figure}

Exponential decay is in line with the expectation that establishing and maintaining the synapses in neural connectomes is associated with wiring cost \cite{bullmore2012economy,chen2006wiring,ahn2006wiring}. 
%\hl{The first reference is a review article that mentions different scales. The other two are on cellular-level C. elegans connectome.} 
The simplest way to define the total wiring cost of the connectome is by summing up the Euclidean distances between the pairs of connected neurons.
%One of the consequences of the exponential distance dependence is the nearly optimal placement of the neurons in terms of minimizing the total wiring cost, similar to Ref. \cite{ercsey2013predictive}. 
To assess the optimality of neuron placement without disrupting the network structure, we randomly shuffle the node positions while keeping the network topology fully intact.
The wiring length obtained from this randomization of the mouse connectome is about 3 times larger than the true wiring length ($p<0.001$, see \cref{fig: shuffle} for the distribution of wiring lengths over 200,000 reshuffled samples). Similarly, the wiring length of the shuffled networks 
%in fly and human is significantly larger than the one observed in the empirical connectomes ($p<0.001$)--- 
is about 2 times the true wiring length in the fly and more than 6 times the true wiring length in the human ($p<0.001$), as illustrated in \cref{fig: shuffle}. Therefore, we conclude that the observed wiring length is much shorter than expected by chance, given the connectome topology.

 We find that the connectome degree sequence alone (model \textbf{k}, sometimes referred to as soft configuration model, shown in red in \cref{fig: randomizations} and discussed in \cref{subsec: ME}) does not imply the observed distance dependence. For instance, in model  \textbf{k}, the average distance between the neighboring mouse neurons is 2.6 times 
larger than that 
of the mouse connectome. This means that, unlike the empirical connectome, model \textbf{k} is not as optimal in terms of the wiring cost---an observation that also holds for fly and human.

%The distance dependence in connectomes is far from random---i.e., $0.34\%$ connection probability for any pair of neurons, independent of distance, expected from the Erd\H{o}s-R\'enyi (ER) model preserving the total number of edges of the mouse connectome. 
%Still, 
Next, we demonstrate that distance dependence alone is insufficient to predict the connectome structure.
For instance, the observed degree distribution (\cref{fig: randomizations} D-F, dark blue) is much broader than that coming from a distance-based model (model \textbf{d}) for all three organisms (\cref{fig: randomizations} D-F, light blue). As an illustration, the interquartile range (IQR) 
%, a measure of the spread of the data) 
of the degree distribution in the mouse connectome is significantly larger than the IQR in model \textbf{d}, 32.0 vs 20.9 ($p<0.001$). Additionally, the mean to median degree ratio is 1.48 in the mouse connectome, compared to the much lower value of .95 in model \textbf{d} ($p<0.001$). This indicates that the size of \textit{hubs}---network nodes whose degree greatly exceeds the average---in the data is not fully captured by model \textbf{d}. 
For example, the number of partners of the highest degree hub---881 in the mouse, which constitutes 13\% of the nodes in the network---is about an order of magnitude larger than that expected based on model \textbf{d}. Similarly, the hub degrees (see \cref{table: basic info} for hub sizes in empirical connectomes) are significantly underestimated by model \textbf{d} in fly and human, as shown in \cref{fig: randomizations} D-F.

Altogether, our results indicate competing implications of the broad degree distribution and the exponential distance dependence. 
To construct a network model that resolves this conflict, in model \textbf{k+L}, we preserve the total distance $L$ between the pairs of connected neurons---%a proxy for 
the overall wiring cost---in addition to maintaining the degree sequence \cite{halu2014emergence}, as discussed in \cref{subsec: exponetial distance preserving}. Unlike model \textbf{k}, model \textbf{k+L} leads to distance dependence that is qualitatively similar to what we observe in data (see top row of \cref{fig: randomizations}, purple line), %which is 
especially %evident 
for the mouse. Note that model \textbf{k+L} requires estimating a distance parameter $d_0^{\textbf{k+L}}$ from data. Notably, the values of $d_0^{\textbf{k+L}}$ are similar for fly, mouse, and human, see \cref{subsec: exponetial distance preserving}, when measured in soma size units, reinforces the finding that some size sets a natural length scale for neural connectome models.

\begin{figure}
	\centering
	\includegraphics[width=1.\linewidth]{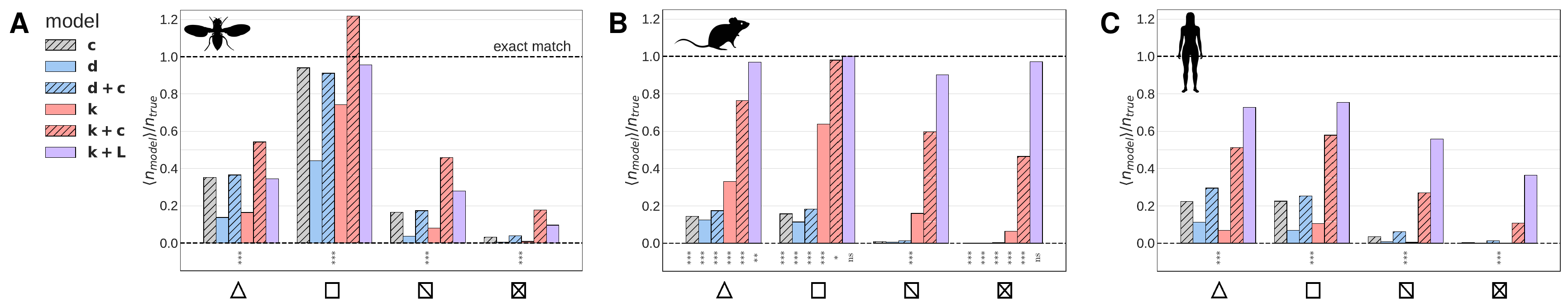}
	\caption{
		Comparing the graphlet counts in empirical connectomes and models based on 100 model realizations.
		Bar plots show the inverse fold changes ($\left< n_{model}\right>/n_{true}$) for the counts of the number of undirected triangle and square graphlets. The labels at the bottom correspond to the two-sided $p$-values obtained from $z$-scores. $n.s.$ (not statistically significant) corresponds to $p>0.05$. $^*$, $^{**}$, $^{***}$ correspond to $p<0.05,~.01$, and $0.001$, respectively.}
	\label{fig: fold changes}
\end{figure}

As the next step, we consider the extent to which models \textbf{d}, \textbf{k}, and \textbf{k+L} generalize beyond the network features explicitly built into them. For example, we consider the counts of different \textit{graphlets}---small connected subgraphs---of the connectome in data and models \cite{milo2002network,prvzulj2007biological}. 
Graphlet counts represent the higher-order local network structure beyond degree distribution, thus providing a sensitive measure of similarity between models and data \cite{prvzulj2007biological}. In \cref{fig: fold changes}, we illustrate the four small undirected graphlets we consider---triangle, square, square with a diagonal edge, and a four-node all-to-all connected graph.
As triangle counts indicate if the neighbors of connected neurons tend to be connected,
%. Since correlated gene expression decays exponentially with distance \cite{fornito2019bridging}, 
an abundance of triangles can hint at similarity-driven connectivity. Similarly, an overrepresentation of squares with no diagonals suggests complementarity-driven wiring rules \cite{kovacs2019network}.
%\ik{Is the genetic rule mentioned here earlier at the same distance scale? If not, then it is not relevant.}
In general, the graphlet counts are better matched by the \textbf{k+L} model (violet), than by models \textbf{d} and \textbf{k} (unshaded blue and red, respectively) (\cref{fig: fold changes}). 

In addition to graphlet patterns, we consider other local and global network measures commonly used in network neuroscience \cite{rubinov2010complex} that we summarize in \cref{subsec: network measures}.
Global measures such as the network diameter and average shortest path length are well-captured by model \textbf{k+L} across the board, with model \textbf{d} and even model \textbf{k} often closely capturing these properties as well, as demonstrated in \cref{fig: fold changes extra}.
However, the advantage of incorporating a combination of network topology and spatial structure in model \textbf{k+L} is most evident for local network measures such as local efficiency, transitivity, and clustering coefficient. 
For instance, the average local efficiency---a measure of how fault tolerant the system is with respect to individual node removal \cite{latora2001efficient}---is relatively low in models \textbf{k} and \textbf{d}, but matches the empirical connectome better in case of model \textbf{k+L}, demonstrating an especially good agreement for the mouse. In contrast, a global analog of this efficiency measure is well-captured and even overestimated by models \textbf{k} and \textbf{d}.
%\ik{Need an extra sentence on why? How do we see this? Like: As we match these measures more closely than other models...}

\begin{figure}
    \centering   \includegraphics[width=.65\linewidth]{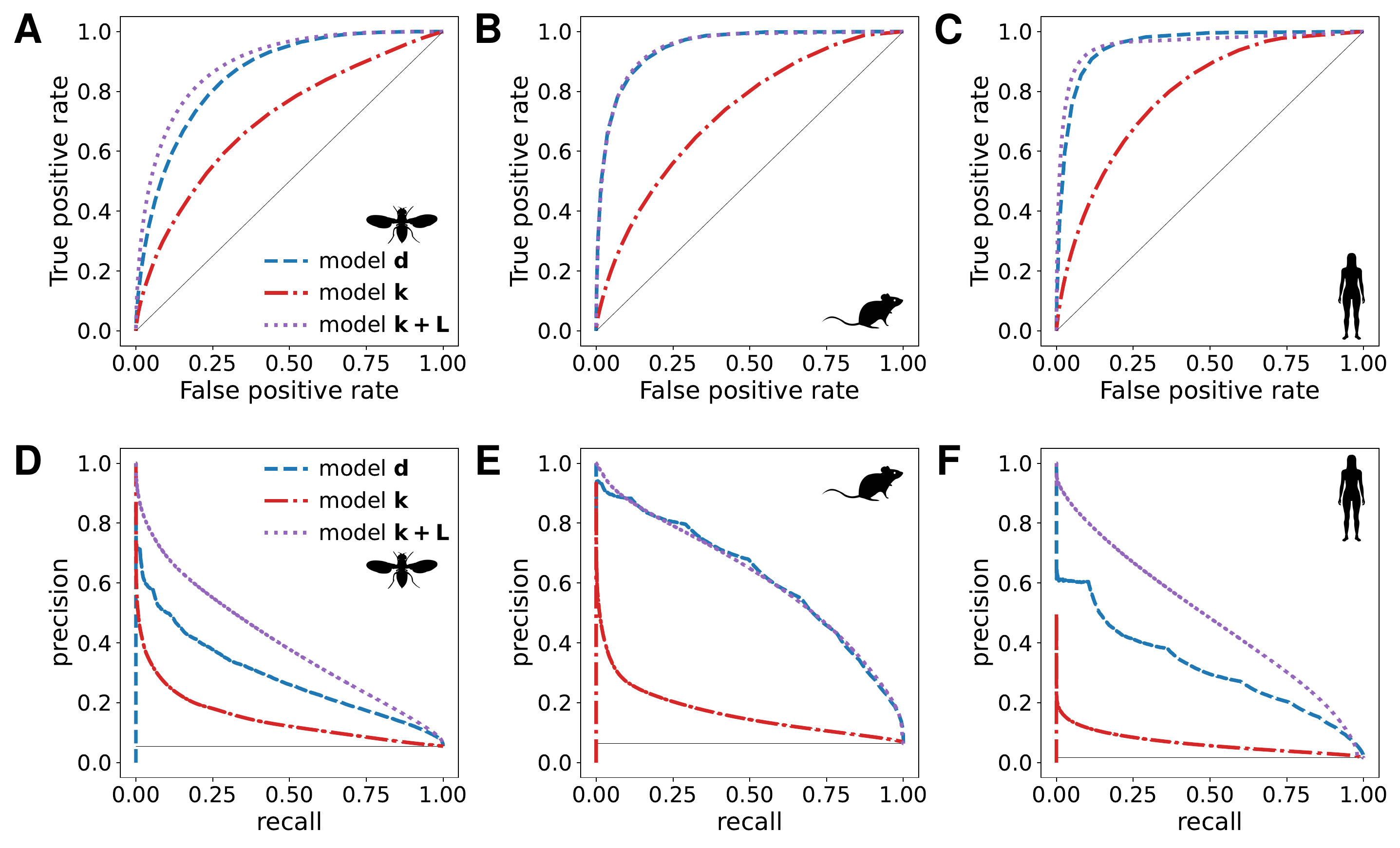}
    \caption{Predicting connectome edges from connectome models. \textbf{A-C:} ROC (receiver operating characteristic) curves for model \textbf{d} (blue), model \textbf{k} (red) and model \textbf{k+L} (purple). \textbf{D-F:} precision-recall curves for the same models. The areas under the curve (AUC) for the three models are provided in \cref{table: AUC}}
    \label{fig: ROC and PR syn}
\end{figure}

%\ik{First discuss predicting the connectome from the connectome:}
So far, we discussed how well the connectome models reproduce the empirical network structure and spatial organization. Another relevant question is the extent to which the models \textbf{d}, \textbf{k}, and \textbf{k+L} capture the existence of individual synapses---a question relevant to network link prediction. In other words, what is the performance of our models in a binary classification task of predicting whether two neurons form a synapse?
To address this question, we first present the Receiver-Operating Characteristic (ROC) curve, which demonstrates the relationship between the true and false positive rates, in \cref{fig: ROC and PR syn} A-C. Additionally, we provide the precision-recall curve for this classification task, as shown in \cref{fig: ROC and PR syn} D-F. This curve is a useful tool to assess the performance of our models given the severe data imbalance---synapses are only present between a small fraction of the available pairs of neurons. A higher area under the curve corresponds to better performance in the classification task for both the ROC and precision-recall curves.
%Additionally, we consider how well the connectome models \textbf{d}, \textbf{k}, and \textbf{k+L} predict connectome edges. 
%The relevant precision-recall and ROC curves are provided in \cref{fig: ROC and PR syn}. 
As demonstrated in \cref{table: AUC}, the best performance is achieved in model \textbf{k+L} across datasets. Similarly, the likelihood of the observed connectome---as calculated using the corresponding maximum entropy model parameters---is the highest for model \textbf{k+L}, see \cref{table: likelihood}.
%\ik{Define the predictions task, ROC, AUC, etc.}

Models \textbf{d}, \textbf{k}, and \textbf{k+L} do not explicitly preserve the physical contact constraint.  
For the mouse, 18\%, 51\%, and 64\% of edges created by models \textbf{d}, \textbf{k}, and \textbf{k+L} respectively are %expected to be 
contained in the contactome. The edge overlap arising from these models is larger than expected from the ER model---e.g., 6.4\% for the mouse. Similarly, model \textbf{k+L} preserves the contact constraint to a larger extent than other models in mouse and human, but the average fraction of overlap varies across the contactomes---being 32\% for the fly, and 53\% for the human. 

\begin{figure}
    \centering   \includegraphics[width=.65\linewidth]{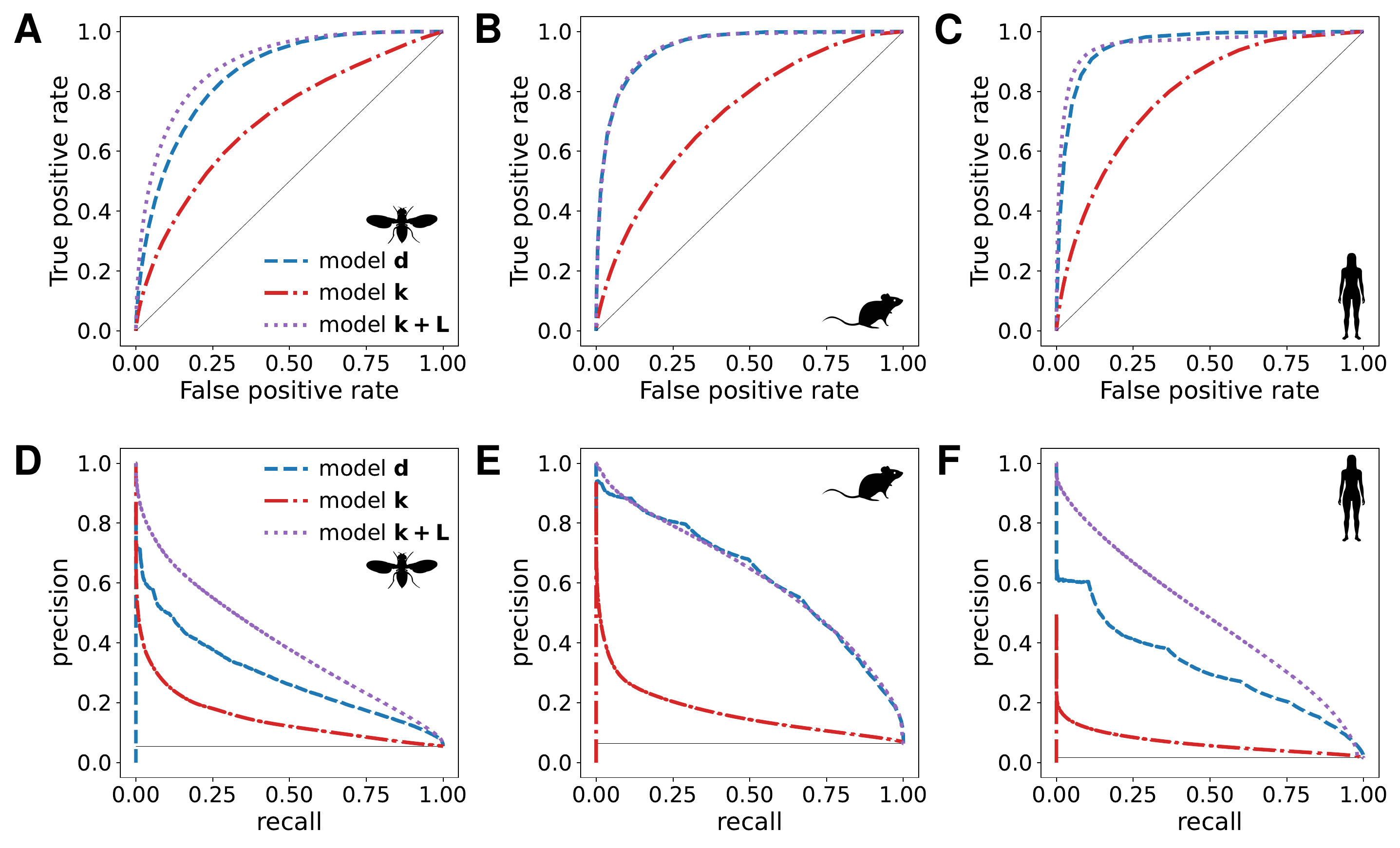}
    \caption{Predicting contactome edges from connectome models. \textbf{A-C:} ROC (receiver operating characteristic) curves for model \textbf{d} (blue), model \textbf{k} (red) and model \textbf{k+L} (purple). \textbf{D-F:} precision-recall curves for the same models. The areas under the curve (AUC) for the three models are provided in \cref{table: AUC}}
    \label{fig: ROC and PR}
\end{figure}

To further assess the accuracy of models \textbf{d}, \textbf{k}, and \textbf{k+L} in capturing the contactome edges, we consider
how well these models perform on a binary classification task that determines whether two neurons are in physical contact.
We demonstrate the ROC curves and the precision-recall curves for this edge classification task %\ik{introduce the task first explicitly}, 
in \cref{fig: ROC and PR}, with the area under the curve (AUC) values provided in \cref{table: AUC}. Model \textbf{k} performs better than chance, but worse than models \textbf{d} and \textbf{k+L}. For the mouse dataset, models \textbf{d} and \textbf{k+L} demonstrate similar performance in retrieving the contactome edges---AUC-ROC is 0.95, and AUC is 0.62 for the precision-recall curve---while for human and fly, model \textbf{k+L} clearly performs best.
%, as especially evident from the precision-recall curves in \cref{fig: ROC and PR}.
As we show in \cref{subsec: contact constraint}, the maximum entropy framework also allows us to explicitly include the contact constraint to enforce that the contactome is a sub-network of the connectome.

So far, we have shown that maximum entropy models with minimal inputs generalize to the properties of binary networks that are not explicitly included in the model. In addition to that, our models are also able to capture the ``edge weights''---as defined by the number of synapses between pairs of neurons---similarly to the models in Ref.~\cite{haber2023structure}. This is evident from Spearman's rank correlation coefficients between edge probabilities $p_{ij}$ and edge weights $w_{ij}$, shown in \cref{fig: weight correlation} A-C. 
%\ik{<- Is this not restricted to the contactome?} 
Models $\textbf{d}$, $\textbf{k}$, and $\textbf{k+L}$ show positive statistically significant ($p<0.001$) correlation with the empirical edge weights across datasets.
%, with model \textbf{k+L} exhibiting the highest correlation across organisms. 
When restricted to $p_{ij}$ among the neural connectome edges (\cref{fig: weight correlation} B), model $\textbf{k+L}$ shows the highest correlation with edge weights in fly, mouse, and human among the three models.
%($p<0.001$ for all models).

\subsection{Modeling the connectome with contact constraints}\label{subsec: contact constraint}

%Close physical proximity (``contact'') between the surfaces of neurons is necessary for synapse formation \cite{peters1976projection,braitenberg1991peters,rees2017weighing,stepanyants2005neurogeometry, reimann2015algorithm}. 
%To capture this constraint, we constructed the physical contact networks---\textit{contactomes}---for the three datasets, see Methods \ref{subsec: contact} \ik{This sounds redundant}. 

Here, we explore the maximum entropy models of connectomes that explicitly include the necessity of physical contact between neurons---as encapsulated by the contactome network---as a hard constraint. The structure of a subset of the multilayer human connectome-contactome network is illustrated in \cref{fig: multiplex network} A. To put that constraint in context, we note that the number of edges in the contactome $E_{\text{cont}}$ is much larger than that in the connectome, $E_{\text{syn}}$---specifically, $E_{\text{cont}}/E_{\text{syn}}\approx4$, $12$, and $27$ for fly, mouse and human. Still, contactome edges comprise less than $2\%$ of all possible neuron pairs, see \cref{table: basic info}, thus significantly narrowing down the space of potential synaptic connections.
%\ik{We had more detailed info on this earlier - maybe move/remove}.
%Thus, physical contact alone does not determine which edges are present in the connectome \ik{We already established this}. 

To uncover the impact of physicality \textit{alone} on the connectome in more detail \cite{posfai2023impact}, we consider random sub-networks of the contactome that preserve the number of edges in the connectome as a soft constraint (model \textbf{c}). This model does not explicitly preserve the distance dependence, wiring cost, or degree sequence. The average distance between neighboring neurons in model \textbf{c} is only 1.12 times larger than the true connectome wiring distance for the mouse. That indicates an improvement in capturing the wiring cost optimization compared to model \textbf{k}---another model that does not explicitly preserve distance dependence. The degree distribution of model \textbf{c} is still less broad (IQR of 29.0 for the mouse) and less dominated by hubs (mean to median degree ratio of 1.0 for the mouse) compared to the empirical connectome (IQR of 32.0, mean to median degree ratio of 1.48 for the mouse).
This is an improvement compared to model \textbf{d}---another model that does not explicitly preserve the degree sequence. However, model \textbf{c} still underestimates the largest hubs. 

The node degrees in the connectome and contactome ($k_\text{syn}$ and $k_{\text{cont}}$, $k_\text{syn} \leq k_{\text{cont}}$, shown in \cref{fig: multiplex network} C insets) are not independent.
The Spearman's rank correlation coefficient is large and positive for the three organisms: e.g., its value is 0.87 for the mouse ($p<0.001$), which shows that the neurons with high contactome degree $k_{\text{cont}}$ tend to have high connectome degree $k_{\text{syn}}$.
However, the relationship between $k_{\text{cont}}$ and $k_{\text{syn}}$ appears to be superlinear (see the insets in \cref{fig: multiplex network})
%More specifically, the relationship is superlinear
---the nodes with many contact partners have a disproportionately large number of synaptic partners.
Again, this demonstrates that the connectome is not a random subnetwork of the contactome, as in the random case, we would observe a linear relation.

\begin{figure}
	\centering
	\includegraphics[width=.9\linewidth]{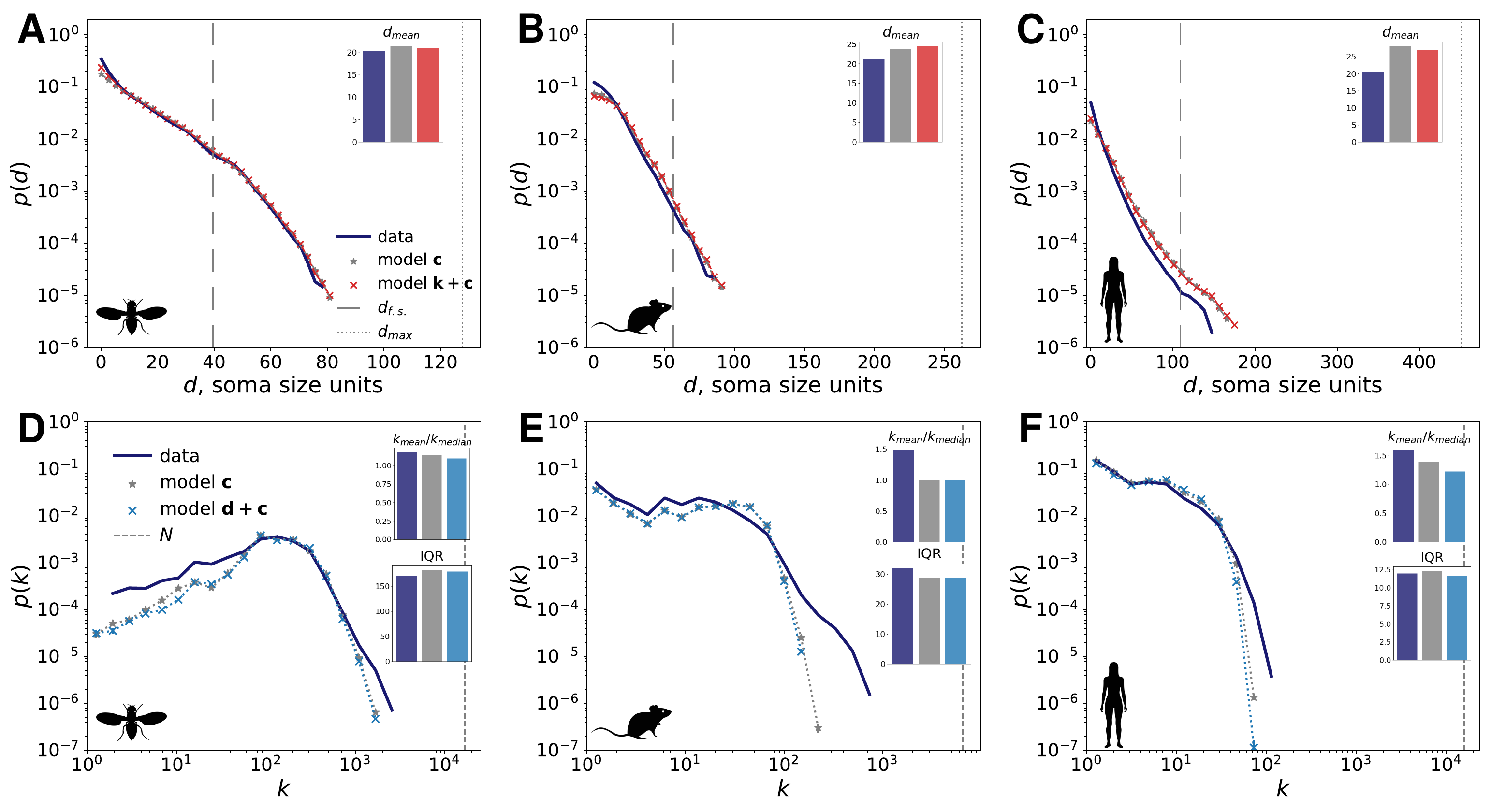}
	\caption{Distance and degree distribution of connectome data and models with contact constraints. Each model distribution shows the average between 100 realizations. \textbf{A-C}: distance distribution of the synaptic network data (dark blue), model \textbf{c} (grey) and model \textbf{k+c} (red). The vertical line at $d_{\text{max}}$ corresponds to the largest distance between the neurons in the dataset. The vertical line at $d_{\text{f.s.}}$ corresponds to the distance between neurons above which the finite size of the experimental volume affects the distance dependence. This distance corresponds to the peaks of the pairwise neuron distance distributions shown in grey in \cref{fig: distance densities}. \textbf{A-C insets}: mean interneuron distance between the connected pairs in data ans models.
		\textbf{D-F}: 
		degree distribution of the synaptic network data (dark blue), model \textbf{c} (grey) and model \textbf{d+c} (light blue).
		\textbf{D-F, upper insets}: ratio of mean degree to median degree in data and model \textbf{d}.
		\textbf{D-F, lower insets}: interquartile range of the degree distribution of data and model \textbf{d}.}
	\label{fig: randomizations contact}
\end{figure}

%Next, we consider the contactome distance dependence. 
Yet, we find that the distance dependence in synaptic networks is largely captured by the contactome in all three organisms. As an illustration, the probability of a connectome edge given the contactome edge between a pair of neurons exists (olive line in \cref{fig: randomizations contact} B) appears largely independent of distance. The quantitative agreement between the connectome and contactome distance dependence can be assessed by comparing the characteristic distance $d_0$ assuming exponential decay $p(d)=\alpha e^{-d/d_0}$---we estimate $d_0$ to be 10 and 12 soma sizes in mouse connectome and contactome. A similar level of agreement in $d_0$ estimates is observed in the fly and human (see the values provided in \cref{table: orientation}). To summarise, the distance dependence of connectomes follows the same trends as that of contactomes. As a result, model \textbf{d+c}---a model that preserves the distance dependence while respecting the contact constraint---does not perform better than model \textbf{c} in terms of capturing the degree distribution of the connectome, see \cref{fig: randomizations contact}. Models \textbf{c} and \textbf{d+c} also underestimate %perform in capturing 
the graphlet counts, as demonstrated in \cref{fig: fold changes}.

Next, we introduce the model \textbf{k+c} that combines topological---in the form of degree distribution---and physical contact constraints. The average distance between the neighboring neurons in this model is 1.15 larger than that in the empirical mouse connectome. Thus, model \textbf{k+c} does not capture wiring cost constraints. The mismatch in wiring cost is significantly ($p<0.001$) larger than that in model \textbf{c}---therefore, there is a trade-off between accurately capturing the wiring cost and preserving the degree sequence. However, model \textbf{k+c} improves on models \textbf{k} and \textbf{c} by better representing the higher-order connectome structure in fly, mouse, and human, as demonstrated by the graphlet counts in \cref{fig: fold changes}. Similarly, model \textbf{k+c} is more accurate at reproducing other local connectome properties, such as local efficiency, transitivity, and clustering coefficient, as shown in \cref{fig: fold changes extra}.

\begin{figure}
    \centering   \includegraphics[width=.65\linewidth]{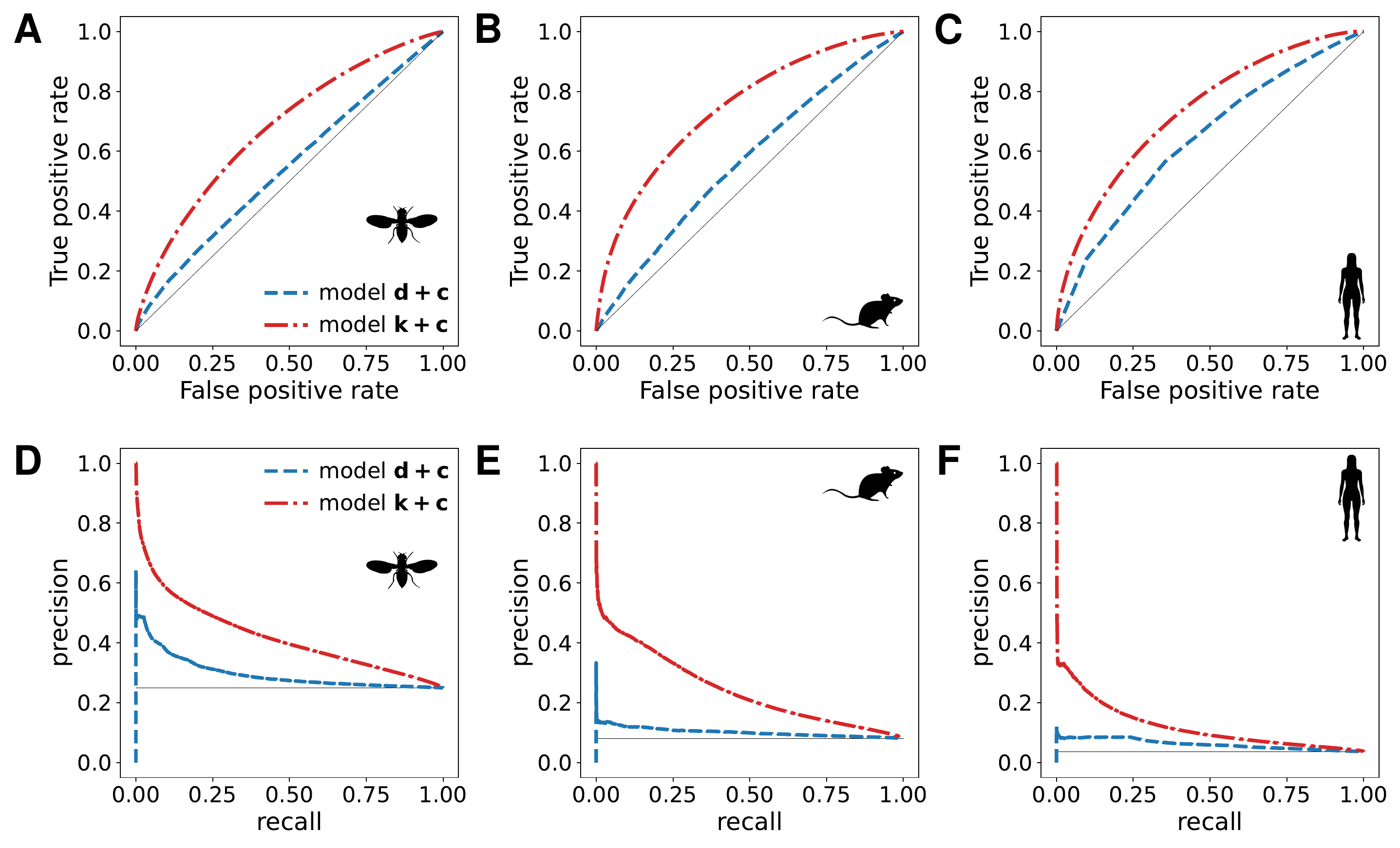}
    \caption{Predicting connectome edges from connectome models with contactome constraints. \textbf{A-C:} ROC (receiver operating characteristic) curves for model \textbf{d+c} (blue), model \textbf{k+c} (red). \textbf{D-F:} precision-recall curves for the same models. The areas under the curve (AUC) for the three models are provided in \cref{table: AUC}}
    \label{fig: ROC and PR syn cont}
\end{figure}

As an alternative way to assess the performance of models \textbf{d+c} and \textbf{k+c}, we determine whether they can predict which contactome edge is also present in the connectome. We again consider the ROC and precision-recall curves (see \cref{fig: ROC and PR syn cont}), together with their AUC values (demonstrated in \cref{table: AUC}). We find that model \textbf{k+c} performs better in this classification task, as indicated by the AUC values. 
%\ik{Why? In what sense? Are all performance measure higher? If so, say so.} 
Model \textbf{k+c} also yields a higher likelihood of the empirical network than models \textbf{c} and \textbf{d+c}, see \cref{table: likelihood}.

\begin{figure}
    \centering   \includegraphics[width=.8\linewidth]{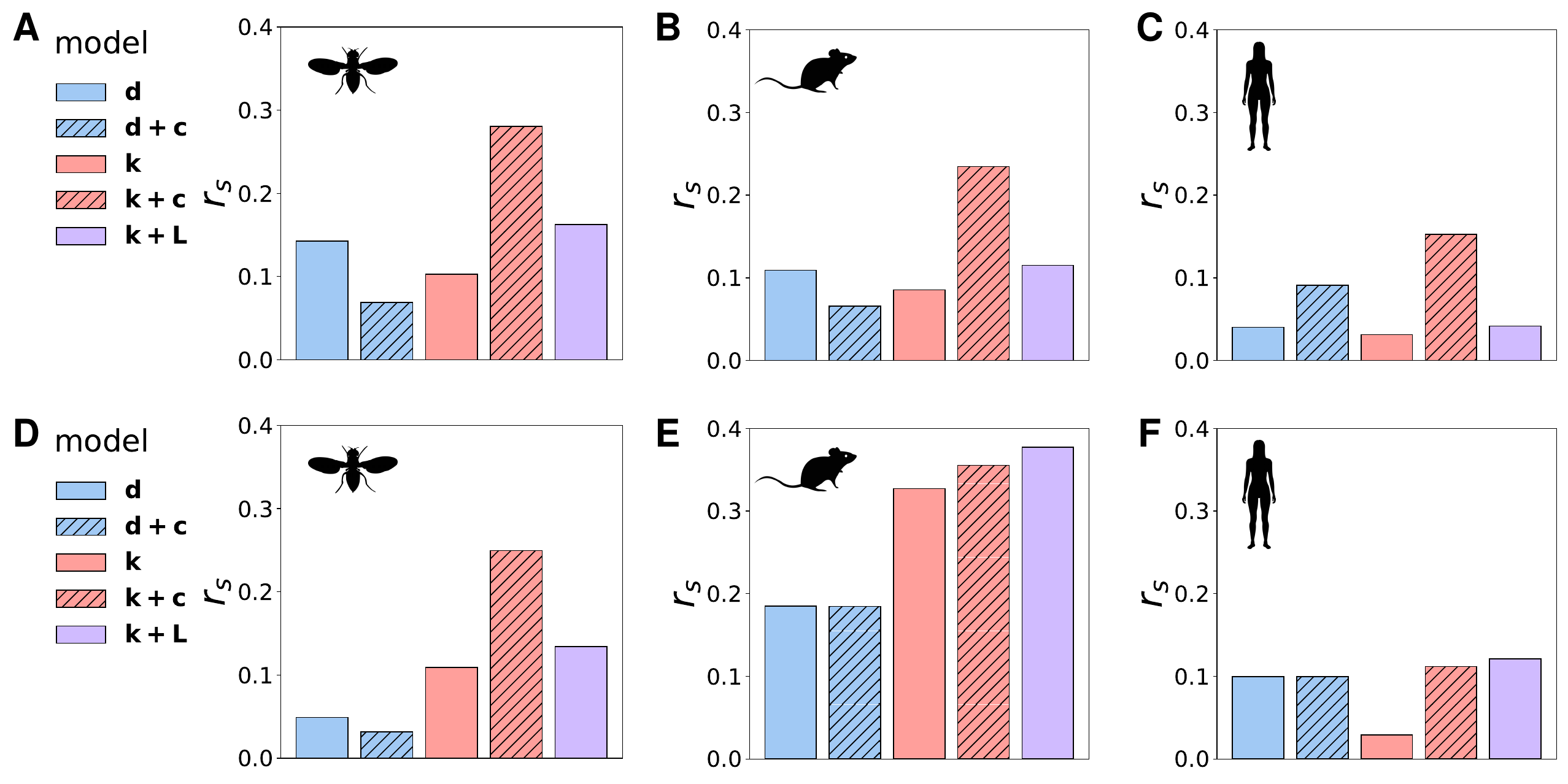}
    \caption{Spearman correlation coefficient between the edge probability predicted by the maximum entropy models and edge weight in data. \textbf{A-C}: correlation calculated for all pairs of nodes (models \textbf{d}, \textbf{k}, and \textbf{k+L}) and pairs of nodes that form contactome edges (models \textbf{d+c} and \textbf{k+c}). \textbf{D-F}: correlation calculated for pairs of nodes that form connectome edges.}
    \label{fig: weight correlation}
\end{figure}

Similarly to models $\textbf{d}$, $\textbf{k}$, and $\textbf{k+L}$ in \cref{subsec: null}, contact-based models partially capture the edge weights (number of synapses between pairs of neurons), even though no weight information is built into the models. Edge probabilities in model $\textbf{k+c}$ show stronger Spearman's rank correlation $r_s$ with the number of synapses than the edge probabilities in $\textbf{d+c}$ for all three organisms. For example, within the mouse connectome edges, $r_s=0.18$ for model $\textbf{d+c}$ and $0.36$ for model $\textbf{k+c}$ ($p<0.001$ in both cases), as shown in \cref{fig: weight correlation} E. In general, models \textbf{k+c} and \textbf{k+L} are consistently the top models in capturing edge weights for the three organisms (see \cref{fig: weight correlation} A-F). Therefore, we conclude that
among the considered models, models $\textbf{k+c}$ and $\textbf{k+L}$ are the best-performing models of the connectome across metrics. Both models reproduce the correct form of distance dependence, while exponential distance dependence is not explicitly built in. 

\begin{figure}
    \centering   \includegraphics[width=\linewidth]{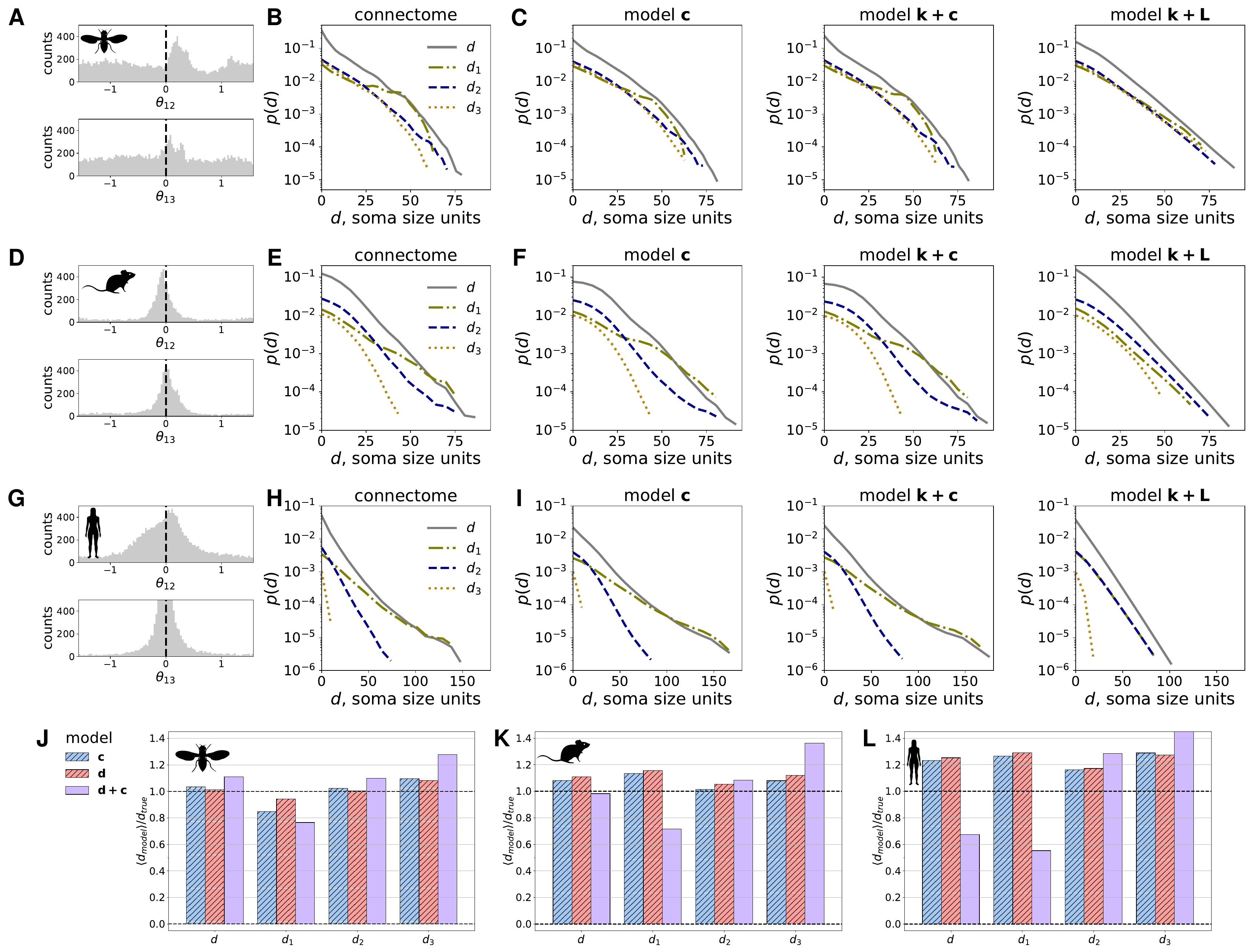}
    \caption{Heterogeneity and neuron orientation dependence in distance dependence for fly (first row), mouse (second row) and human (third row). \textbf{A, D, G}: distribution of angles $\theta_{1i}$ between the dominant PCA vectors of individual neuron meshes and $pc_1$, the dominant principal component obtained from calculating the PCA of all the individual neuron orientations in the $pc_1$-$pc_2$ plane (top) and $pc_1$-$pc_3$ plane (bottom). See \cref{table: orientation} for the PCA vector values and more details in the caption. \textbf{B, E, H}: 3D Euclidean distance dependence (grey) and the distance dependence in $pc_1$, $pc_2$, and $pc_3$ directions (olive, dark blue, and dark yellow) in the connectome. \textbf{C, F, I}: distance dependence for models \textbf{c}, \textbf{k+c}, and \textbf{k+L}. \textbf{J-L}: inverse fold changes for the characteristic distances in 3D and along the principal components. See \cref{table: orientation} for the $d_i$ values.}
    \label{fig: neuron orientation fold change}
\end{figure}

However, distance dependence in neural connectomes is not fully homogeneous. For instance, the mouse and human connectomes we use come from cortical datasets, each containing several cortical layers \cite{bae2021functional,shapson2021connectomic}, and the distance dependence within the layers can be different from distance dependence across the layers along the cortical columns. To treat the three datasets in a consistent way, we find the principal component vectors corresponding to the neuron alignment, see \cref{table: orientation}, \cref{fig: neuron orientation fold change}, and their captions. The details of distance dependence $p(d_i')$, where $d_i'$ is the 1D distance along a specified component, differ across principal components ($i=1,2,3$) (see \cref{fig: neuron orientation fold change}, panels B, E, H). Assuming $p(d_i')\propto e^{-d_i'/d_i}$, by construction it is true that $d_1>d_2>d_3$. For example, we find $d_1=15$, $d_2=10$, and $d_3=8$ in the mouse in soma size units (see \cref{table: orientation}). Likewise, we estimate the orientation-dependent distance scale in fly and human. In the human data, we find similar heterogeneity in $d_1$ and $d_2$---with a caveat that $d_3$ is very small since the human brain sample is comparatively very thin. In the fly data---where, unlike in cortical layers of mouse and human, the neuron orientation is more radial, see \cref{fig: soma vs cm distance} A---we also observe that distance dependence varies across the principal component directions. In the fruit fly, in contrast to other datasets, $d_1$ appears to have a different functional form from $d_2$ and $d_3$.

Last, we compare the performance of the considered models on capturing the heterogeneity of distance dependence in empirical connectomes. As shown in \cref{fig: neuron orientation fold change} panels C, F, and I, contactome (model $\textbf{c}$) qualitatively exhibits similar orientation dependence as the connectome, and the orientation dependence is also evident in model $\textbf{k+c}$. On the other hand, model $\textbf{k+L}$ does not capture the heterogeneities in orientation dependence. Quantitatively, this is demonstrated in panels J-L of \cref{fig: neuron orientation fold change}. Thus, using the contactome constraint facilitates capturing the details of distance dependence, which are not captured even by model $\textbf{k+L}$ that performs well based on other metrics.

\section{Discussion}

Connectomes are shaped by an interplay between spatial constraints and the need for complex topology that is essential to brain function \cite{bullmore2012economy}.
To quantify the shared spatial properties of connectomes, we establish that distance dependence in undirected unweighted millimeter-scale neural connectomes is exponential. We also demonstrate that the characteristic distance scales are similar across species when expressed in soma-size units.
%that we derived from volumetric data
Moreover, we show that the neuron positions are highly optimal in terms of minimizing the wiring cost. 
Driven by a biological hypothesis, we construct physical contact networks from volumetric data and show that exponential decay follows from the contactome structure. 
Similarly, the contactome captures the orientation dependence in the connectome.
%Similarly, the heterogeneity of distance dependence in the connectome is captured by the contactome. 
Thus, establishing 
accurate spatial or physical \cite{pete2023network} generative models of the contactome can help elucidate the origin of the exponential distance dependence across species.  

Recently, signatures of structural criticality---with critical exponents consistent across organisms---were observed in the three volumetric brain datasets we analyze here \cite{ansell2023unveiling}. Such universal properties of the %physical structure of the neurons and their spatial arrangement 
brain anatomy are expected to have a profound impact on the spatial and topological organization of the contactome---and, consequently, connectome---and lead to similarities across species. Models of brain anatomy that belong to the same universality class also lead to models of the contactome, a direct input for some of our models. At this point, it remains an intriguing open question to determine whether structural criticality is required for the complex contactome structure presented here, and how it relates to brain function.

The contactomes we constructed for three organisms are useful beyond serving as spatial constraints for the connectome models. The contactomes can aid in predicting still missing chemical synapses in the reconstructions.
In addition to chemical synapses, contactomes must contain electrical synapses---also know as gap junctions---that are much smaller and therefore hard to detect in raw volumetric data \cite{maeda2009structure}.
Thus, contactomes could serve as a useful constraint in gap junction prediction as well \cite{kovacs2020uncovering}. 
As detailed and accurate classification of neurites into axons and dendrites becomes routinely available in high-resolution volumetric datasets \cite{shapson2021connectomic}, contactomes can become useful in analyzing the effects of subcellular wiring specificity on the connectome structure. Additionally, as directly comparing individual connectomes within the same species is already becoming feasible for complex organisms such as the fruit fly \cite{schlegel2023whole}, the structure of the corresponding contactomes can be used as an extra measure of intraspecies variability. Last but not least, the novel approach of considering the multilayer connectome-contactome network goes beyond the standard spatial network framework and thus could ignite new research in network science. 

As a first step in neural connectome topology analysis, we establish that the degree distribution is broad across datasets, yet not scale-free. However, the degree distribution of the empirical connectomes does not follow from spatial constraints alone. Similarly, distance dependence is not implied by the connectome degree sequence.  However, the models that resolve the apparent conflict by using a combination of spatial constraints and degree sequence---\textbf{k+c} and \textbf{k+L}---capture the network structure across species. These models demonstrate predictive power for connectome properties not explicitly built into them. For instance, the number of synapses---a proxy for edge weight---between the neurons is positively correlated with the edge probabilities implied by our models. 
%Assigning higher probability to edges with larger weights $w$ is a useful byproduct of our models. 
The authors of Ref.~\cite{schlegel2023whole} demonstrate that in the fly brain, the connections with higher weight %($w>10$) 
are significantly more likely to be found across hemispheres and across different organisms than edges with smaller weights---thus, more stereotyped edges are expected to be more likely to be captured by our maximum entropy models.

Note that the details of model performance vary across organisms---consistently with the findings of Ref.~\cite{haber2023structure}. For instance, model \textbf{k+L} appears to be the best model to capture the basic structural features of mouse and human connectomes, while model \textbf{k+c} appears to demonstrate better performance on the fly data. The empirical distance dependence is not explicitly built into models \textbf{k+L} and \textbf{k+c}, but in both cases it is well reproduced across organisms. However, model \textbf{k+L} does not capture the distance dependence heterogeneity that is picked up by model \textbf{k+c}. We conjecture that just like in the fruit fly, model \textbf{k+c} will perform better than model \textbf{k+L} once multiple brain regions are captured by the datasets.
The discrepancies between models \textbf{k+c} and \textbf{k+L} and the empirical connectomes are evident from the statistically significant differences in their graphlet counts and network measures (see \cref{fig: fold changes} and \ref{fig: fold changes extra}. These discrepancies can be informative
of the wiring rules missing from our models, e.g., wiring specificity based on neuron type, gene expression profiles, or the type of neurites in physical contact
\cite{udvary2022impact,kovacs2020uncovering,haber2023structure}.

Exponential distance dependence arises naturally in models \textbf{k+c} and \textbf{k+L}. However, we use the empirical neural connectome degrees to obtain the appropriate neuron-level parameters that we then use in these generative models. At the same time, node degrees---and therefore the node-level model parameters---are non-trivially linked to structural properties of corresponding neurons (e.g., surface area, linear span, and their morphology in general). For instance, larger neurons are capable of forming synapses far away from their soma or center of mesh locations and, therefore, have more neighbors, while the location of positions of neighbors of small neurons is confined to a relatively smaller neighborhood. Linking the node-level parameters of generative models to observable structural and biological properties of neurons is a promising direction for future exploration that could link maximum entropy models to the bottom-up models of brain structure and development \cite{oldham2022modeling,kaiser2004modelling}.

Our approach to analyzing volumetric brain data is in line with comparative connectomics \cite{van2016comparative}---an emerging field that aims to uncover the general principles of brain network architecture and identify species-specific features of connectomes. The models we establish can serve as baselines to compare different neural connectome datasets. For instance, they can be useful in comparing different brain regions within the species, instances of the same brain region across healthy individuals, or differentiating the structure of the brain in healthy and diseased states.

Our analysis sets the stage for a thorough investigation of the existing and forthcoming high-quality volumetric brain reconstructions. The maximum entropy models we introduce are scalable, as they can be accurately solved using a simple iterative procedure \cite{vallarano2021fast}. This property, as well as the ability to directly sample the models without using Markov-chain Monte Carlo sampling, makes applying our methodology feasible even for larger connectomes---for instance, the existing fruit fly brain reconstruction \cite{dorkenwald2023neuronal} or regions of the proposed complete reconstruction of the mouse brain \cite{abbott2020mind}.
In addition, maximum entropy models are flexible and can be easily extended to include additional important features of connectome edges---for instance, directionality, weight (e.g., the number of synapses between the neurons or synapse sizes)\cite{bianconi2009entropy,lynn2024heavy}, sign (excitatory or inhibitory) \cite{hao2023proper,gallo2023strong}, the type of neurites involved in synapse formation, or even multi-way interactions between the neurons \cite{saracco2022entropy,xu2020connectome}---as well as their combinations.
Analyzing the graphlet signatures of these more nuanced models provides an opportunity to gain additional insights into the structure of the connectome, as well as its relation to brain dynamics and function \cite{lizier2023analytic,haber2023structure}. The graphlet analysis itself can also be generalized to explicitly include spatial information \cite{kim2012spatiotemporal} in addition to topology, as well as node and edge labels, representing biologically relevant information. More detailed biological input could capture neuron gene expressions or cell types, or even different types of cell labels, such as neurons and glial cells \cite{fields2015glial}, potentially incorporated into the maximum entropy models to reveal the effect of wiring specificity on connectome structure \cite{haber2023structure,schneider2023cell}.

\begin{acknowledgments}
We thank Helen Ansell for providing the physical boundaries for the volumetric mouse and human datasets and for useful discussions.
\end{acknowledgments}

\section*{Code and data availability} 
A repository containing the pre-processed data, model outputs, and model code is available at \url{https://github.com/asalova/neural-connectome-structure}. 

\section{Methods}

\subsection{Synaptic network construction: topology}\label{subsec: synaptic network}

\textit{General considerations}

To construct a neuronal connectome, we need to define the nodes and edges of the network. In general, we want the nodes to correspond to individual neurons, while the edges should represent the presence of at least one synaptic connection. In practice, the fly, mouse, and human datasets we analyze contain reconstructed segments corresponding to parts of individual cells. Thus, each neuron can be present in more than one segment. To avoid representing a single neuron as multiple nodes, we restrict ourselves to segments that contain a soma in the experimental volume---this ensures each neuron is only counted once. Moreover, we only consider the cells with a \textit{single} identified soma to avoid picking up cell merging errors. We present the basic properties of the connectomes we obtain in \cref{table: basic info}. Below, we discuss the specific steps we took to obtain the fly, mouse, and human neuron connectomes.

\textit{Fly (see Ref. \cite{xu2020connectome})}

We use the neurons that are labeled as ``traced'' and ``uncropped'' and whose soma positions within the experimental volume are available (16,804 neurons). The relevant part of the synaptic network is obtained from the compact connection matrix summary v1.2 release available at \url{https://www.janelia.org/project-team/flyem/hemibrain}. We use the version of the network with all of the detected synapses, which results in a total of 9,123,275 synapses in our connectome.

\textit{Mouse (see Refs. \cite{bae2021functional,elabbady2022quantitative})}

Note that the mouse data we used---including neuron ids, the meshes representing their volumetric structure, and synapse information---was obtained in September 2022. Since then, the datasets have been edited and improved, as discussed in Ref.~\cite{bae2021functional}. %\hl{Something about data availability: we provide the networks we used.}
We start by obtaining soma information from the ``nucleus\_neuron\_svm'' table using CAVE, as discussed in \url{https://github.com/AllenInstitute/MicronsBinder/blob/master/notebooks/mm3_intro/CAVEsetup.ipynb} \cite{elabbady2022quantitative}. We then filter the cells by having exactly one labeled soma. Finally, we only consider the cells labeled as ``neuron''. There are 64,360 cells in the mouse dataset that satisfy these conditions. However, many neurons in the resulting connectome are cropped by the experimental volume boundaries, hindering our ability to assess their size and full spatial extent. To make the network similar to the fly dataset and avoid the effects of cropping the neurons on the spatial and topological properties of the connectome, we identify and use the uncropped neurons.

To find the uncropped neurons, we first define the volume boundary by finding the regions in space where no cells are detected for each slice in $z$ dimension and finding the boundaries of each region using the MATLAB Image Processing Toolbox. Then, we thicken the boundary (the thickness of the boundary we use is larger than the largest distance between mesh vertices for lod=1 to ensure all the boundary crossing neurons are filtered out) and remove any ``holes'' that appear due to misalignment of different imaged $z$ slices. As our next step, we identify the meshes that do not have any vertices on the boundary at lod=1. The resulting 9,118 cells become our uncropped neuron candidates.

\begin{figure}
    \centering
    \includegraphics[width=.8\linewidth]{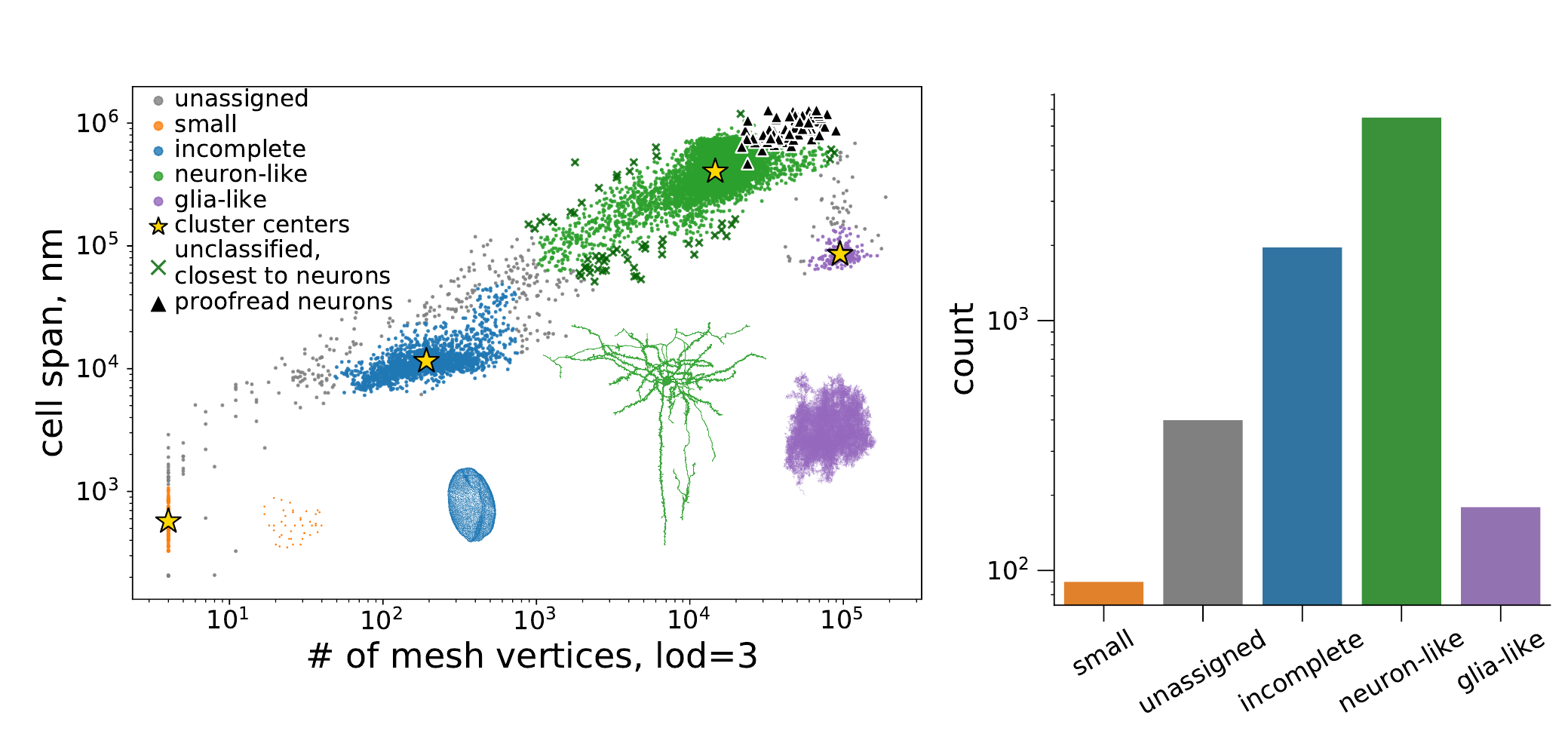}
    \caption{Classification of mouse segments labeled as ``neurons''. Left: distinct clusters of cells (shown in different colors) as identified by DBSCAN based on the``cell span" (cell size in the leading PCA vector direction, where the PCA is performed on the mesh vertices for lod=3) and the number of mesh vertices for the same resolution, together with examples of each type of cell shown in the same color as their clusters. Right: number of cells of each type. }
    \label{fig: mouse classification}
\end{figure}

Some of the segments among the uncropped neuron candidates do not appear to correspond to neurons. Specifically, some cells labeled as ``neurons'' in the dataset appear to be glial cells (e.g., see the purple cell in \cref{fig: mouse classification}, left panel). Other cells appear to only contain a fraction of the cell (e.g., see the blue object that looks like a soma in \cref{fig: mouse classification}, left panel) or arise from other segmentation errors. Plotting two spatial properties of neurons---their span and number of mesh vertices at lod=3 (a proxy for their surface area)---reveals that the neuron candidates form clusters in this 2D space.
To obtain the cluster boundaries numerically, we use the DBSCAN algorithm with parameters $\varepsilon=.1$ and $\text{min}\_\text{samples}=30$ \cite{scikit-learn}.
This leads to five categories (shown in \cref{fig: mouse classification})---two with relatively short span and relatively low number of mesh vertices (``small'' and ``incomplete'' cells), one with a large number of mesh vertices and moderate cell span (``glia-like'' cells), a large cluster containing 6,489 ``neuron-like'' cells, and unclassified cells that were not assigned to any of the clusters. 

To validate our selection of the largest cluster as the one corresponding to neurons, we consider the 78 proofread neurons with extended axons and somas. While none of the proofread neurons are uncropped, their positions on the span vs number of mesh vertices plot (black triangles in \cref{fig: mouse classification}) overlaps with the location of the ``neuron-like'' cluster.
Finally, we use the cells in ``neuron-like'' cluster 
together with 59 additional segments that were originally unclassified but are closest to the center of mass of the ``neuron-like'' cluster as the nodes of the connectome  (6,548 nodes total). 
We obtain the synapses---163,188 in total---associated with these 6,548 neurons from \url{https://bossdb-open-data.s3.amazonaws.com/iarpa_microns/minnie/minnie65/synapse_graph/synapses_pni_2.csv}.

\begin{figure}
    \centering
    \includegraphics[width=.6\linewidth]{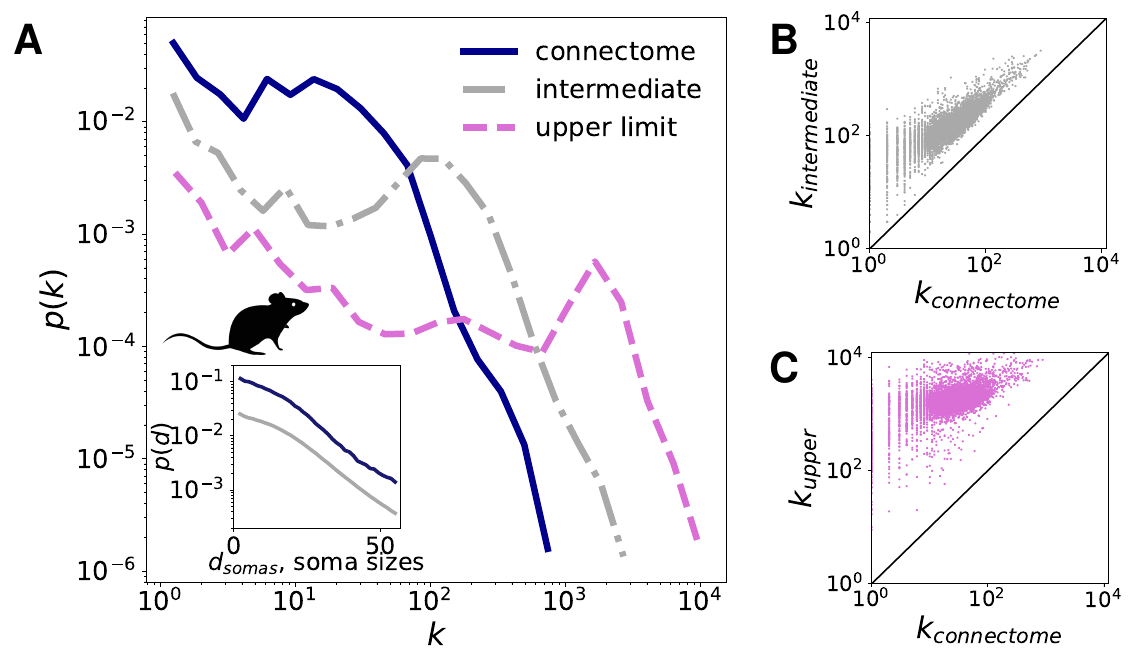}
    \caption{Comparing the node degrees in the mouse connectome we use and its upper bounds. 
    In all cases, the degree distribution is shown for the uncropped neurons as defined in \cref{subsec: synaptic network}. 
    In the degree calculation, the connectome degree distribution (dark blue) accounts for the neighbors that are themselves uncropped neurons.
    The ``intermediate'' degree distribution (grey) includes all neighbors with a single soma in the experimental volume. The ``upper limit'' degree distribution (pink) accounts for all neighbors assigned a unique id. \textbf{A}: degree distributions.  Inset: distance dependence for the connectome and the intermediate network. Here, distance is defined as the Euclidean distance between somas. The estimated exponential scales $d_0$ ($p(d)\propto e^{-d/d_0}$) are similar---11 soma sizes for the connectome and 12 soma sizes for the intermediate network. \textbf{B}: degree correlation between the connectome and intermediate network. Pearson correlation coefficient is $r_p=0.89$. \textbf{C}: degree correlation between the connectome and upper limit network, $r_p=0.55$.} 
    \label{fig: mouse limits}
\end{figure}

Furthermore, we assess the effect of only using the connections between the uncropped neurons on the connectome structure by considering our connectome degrees as a lower limit on node degrees. The ``medium'' and ``upper'' bound degree distributions are provided in \cref{fig: mouse limits} A---both of them are broad and non-power law. The ``intermediate'' degree sequence---where, we include all single-soma neuron neighbors of our fully contained neurons---shows high Pearson correlation with node degrees in our connectome, see \cref{fig: mouse limits} B.

\textit{Human (see Ref. \cite{shapson2021connectomic})}

We obtained the soma labels for individual neurons
from \url{gs://h01-release/data/20210601/c3/tables/somas.csv}. For our analysis, we select the neurons that have a single labeled soma---15,730 cells in total. We acquired the list of synapses from the files in \url{gs://h01-release/data/20210601/c3/synapses/exported/json}. 
The synapse dataset also includes the labels of parts of the neuron involved in synapse formation (e.g., axon to dendrite). However, we do not perform any filtering based on this information and use the entire dataset containing 115,165 synapses. 

The shape of human brain experimental volume is qualitatively different from that of the mouse or fly dataset. Namely, the size of the experimental volume in $z$ dimension is  much smaller than that in $x$ and $y$. We can still identify the neurons that do not cross the experimental volume boundary (967 cells). Alternatively, we can also identify the neurons that do not cross the $xy$ boundary while possible intersecting the boundaries of the experimental volume in $z$ (7,176 cells). However, in both of cases, much of the network structure is lost (330 and 24,144 edges are preserved in a network with neurons fully contained within the $xy$ boundary and the full experimental  boundary respectively) and the network is largely disconnected. Thus, we keep the entire human network in our analysis.

\subsection{Synaptic network construction: spatial properties}\label{subsec: spatial aspects}

\begin{figure}
    \centering
    \includegraphics[width=1.\linewidth]{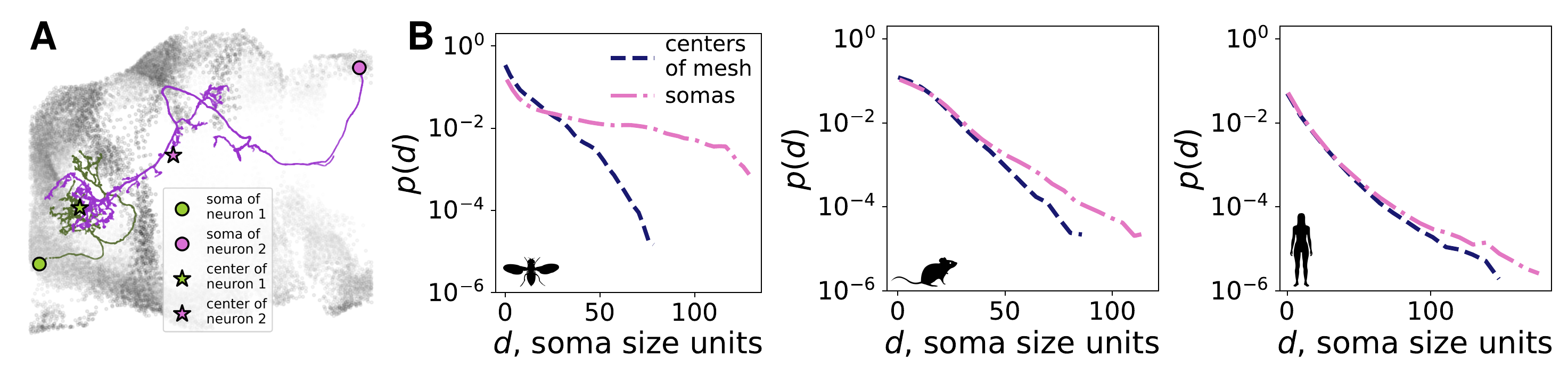} \caption{Distance dependence for different definitions of neuron locations. \textbf{Left}: an example of two neurons (neuron 1 in green, neuron 2 in purple) in the fly brain. The distance between their centers of mesh (shown as stars) is significantly smaller than the distance between their somas (circles). The positions of somas of all the uncropped neurons are shown in light grey. \textbf{Right}: distance dependence using soma locations (purple) vs centers of mesh (dark blue).}
    \label{fig: soma vs cm distance}
\end{figure}

\begin{figure}
    \centering
    \includegraphics[width=1.\linewidth]{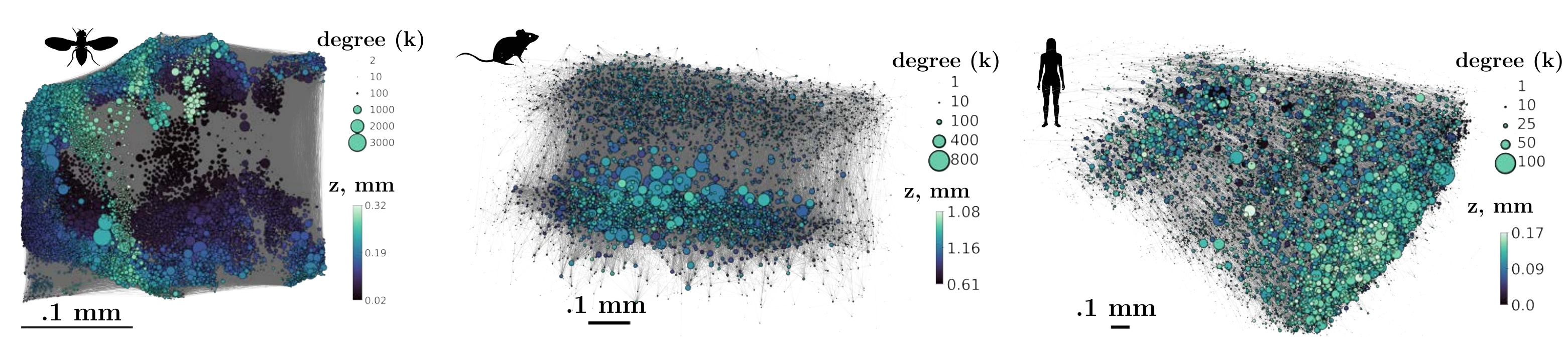}    \caption{Synaptic networks from soma positions. Here, node positions correspond to the $xy$ location of neuron somas, node color and size correspond to their undirected degree and $z$ position respectively.}
    \label{fig: basic properties somas}
\end{figure}

The connectome is inherently spatial. Individual neurons are complex objects embedded in physical space, whose spatial organization relative to other neurons affects who they can form synapses with. The simplest way to capture the neuron location is via its soma position, as illustrated by node positions in \cref{fig: soma vs cm distance}. Then, the relative locations of neurons can be represented by the Euclidean distance between their somas. 
This connectome representation leads to the edge probability that clearly decreases with distance in mouse and human connectomes (see \cref{fig: soma vs cm distance} purple dash-dotted line, see \cref{subsec: soma sizes} for the definition of soma size units). However, the fly edge probability decreases very slowly at larger distances (approximately 20 to 120 soma sizes, as shown in \cref{fig: soma vs cm distance}). This is largely the result of the distinct spatial organization of neurons in the fruit fly brain, where the somas (dots in \cref{fig: soma vs cm distance} A and \cref{fig: basic properties somas}) occupy the periphery of the brain, and the neurites are located closer to the center of the brain (\cref{fig: soma vs cm distance}, see example neurons shown in green and purple) \cite{xu2020connectome}. Thus, the soma position in the fly is not representative of the neuron location as a whole.

To represent the neuron positions more effectively, we use the ``center of mesh''---the center of mass of mesh vertices of individual neurons that we calculated at lod=1 for the three organisms. This center of mesh can be thought of as the center of mass of the \textit{surface} of individual neurons. Defining the distance between neurons as the Euclidean distance between the centers of mesh leads to a more rapid decay of the fly distance dependence and a narrower range of distances between the neurons in synaptic contact (\cref{fig: soma vs cm distance}, dark blue dashed line). To a lesser extent, similar trends are seen in the mouse and human.

\begin{figure}
    \centering
    \includegraphics[width=1.\linewidth]{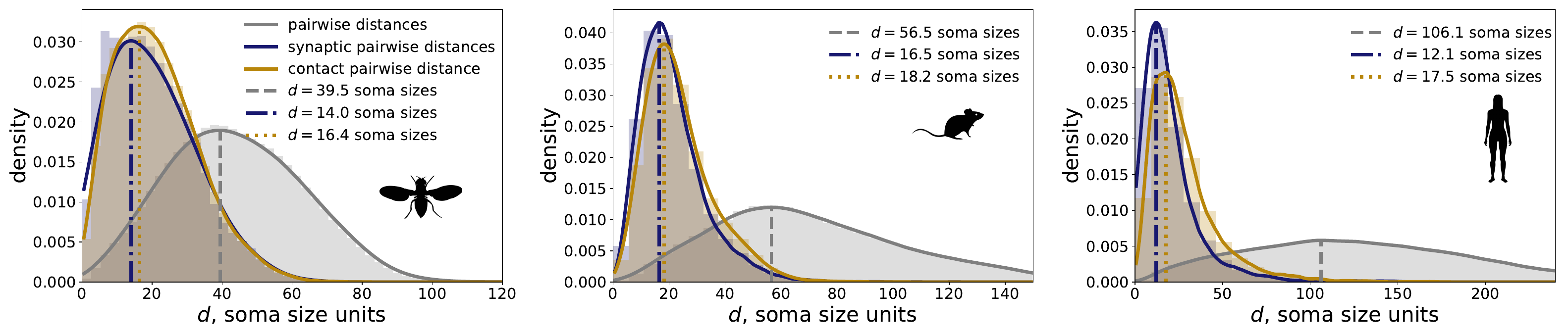}    \caption{Distribution of distances between all pairs of neurons (grey), pairs of neurons connected by connectome edges (dark blue), and pairs of neurons connected by contactome edges (dark yellow). The locations of the peaks of these distributions were estimated using kernel density estimation. The peaks of the synaptic and contact pairwise distances represent the typical distances between the neurons in the connectome and contactome. The peak of the entire distance distribution (grey) represents the distance above which the effects of the finite size of the experimental volume affect the inferred distance dependence.}
    \label{fig: distance densities}
\end{figure}

Note that the spatial connectomes we analyze are cropped---thus, distance dependence for $d$ above a certain threshold does not represent the full neural connectome distance dependence at those distances $d$. To obtain a distance threshold, we consider the pairwise distances between all pairs of neurons for each organism (shown in grey in \cref{fig: distance densities}). The peaks of these distance distributions roughly correspond to the distance threshold we seek---e.g., we estimate it to be $d_{thr}=56$ soma sizes for the mouse.

The spatial connectome we constructed can be used to estimate the distance scale $d_0$ of the exponential decay $p(d)\propto e^{-d/d_0}$ in the three organisms. Using the distance data below $d_{thr}$, we estimate $d_0=10$, $10$, and $12$ soma sizes for fly, mouse, and human respectively---the values are close across organisms. Without the distance threshold, these estimates become $d_0=9$, $9$, and $15$ for fly, mouse, and human. We also expect the distance dependence to not be uniform---e.g., distance dependence along the cortical columns could be different than distance dependence within the cortical layer in mouse and human datasets. This is quantified and illustrated in \cref{table: orientation} and \cref{fig: neuron orientation fold change}.

Finally, we consider the effect of only including the edges between the uncropped neurons in the estimated mouse $d_0$ value. To do so, we compare the distance dependence in the connectome with the distance dependence in a network that includes the edges between the uncropped neurons and cropped neurons with a soma in the volume, see \cref{fig: mouse limits} A inset. Here, we define distance as the Euclidean distance between somas. We get similar values for the two networks---$d_0=11$ and $12$ soma sizes.

\subsection{Soma size estimation}\label{subsec: soma sizes}

We define the soma size as a quantity roughly corresponding to the typical soma radius.
To estimate the fly and mouse soma sizes from data, we collected the distances $d$ from the center of the soma to physical contact locations. Using physical contact locations instead of synapse positions enables using more data points and ensures that the soma size is correctly estimated for the fruit fly, where no synapses are formed directly on the soma \cite{xu2020connectome}. We expect the typical soma size to be located at the peak of the distance distribution (shown as a histogram in grey in \cref{fig: soma sizes}) at low $d$. 

The approach outlined above did not result in a clear peak in the human dataset (grey histogram, right panel in \cref{fig: soma sizes}). 
Fortunately, synaptic contacts in the human dataset are classified based on which parts of the pair of neurons (e.g., axon and soma) were involved in the formation of the synapse \cite{shapson2021connectomic}. Once we restricted the data to only include the synapses connecting axons or dendrites of the pre-synaptic neuron to the soma of the post-synaptic neuron, a peak emerged.

To estimate the soma size $r_{\text{soma}}$ from data, we perform kernel density estimation with linear kernel form and the bandwidth parameter of $0.2$, $0.5$, and $1.5$ for fly, mouse, and human (dark blue line in \cref{fig: soma sizes}), and find the location of the peak of this function ($r_{\text{soma}}\approx 2.47\mu$m, $5.34\mu$m, and $7.6\mu$m for fly, mouse, and human respectively, shown in red in \cref{fig: soma sizes}) \cite{scikit-learn}. The fly soma size obtained using our methodology is in excellent agreement with the average of the soma radii provided in the janelia hemibrain dataset \url{https://www.janelia.org/project-team/flyem/hemibrain} for the neurons we consider (see \cref{subsec: synaptic network}): $2.467\mu$m.

Soma size units define a scale that unifies the three connectomes. For instance, the characteristic distance $d_0$ obtained from the degree and wiring length-preserving maximum entropy model is approximately 9 soma sizes across organisms (see \cref{subsec: exponetial distance preserving}), and the average distance between synaptically connected neurons is approximately 20 soma sizes ($20.27$ for fly, $21.19$ for mouse, and $20.42$ for human). Similarly, the scale of the distance dependence in both connectome and contactome---labeled as $d_{tr}$ and $d_{tr}^c$ in \cref{table: orientation})---is of the same order of magnitude for the three organisms.

\subsection{Contact network construction}\label{subsec: contact}

\begin{figure}
    \centering   \includegraphics[width=\linewidth]{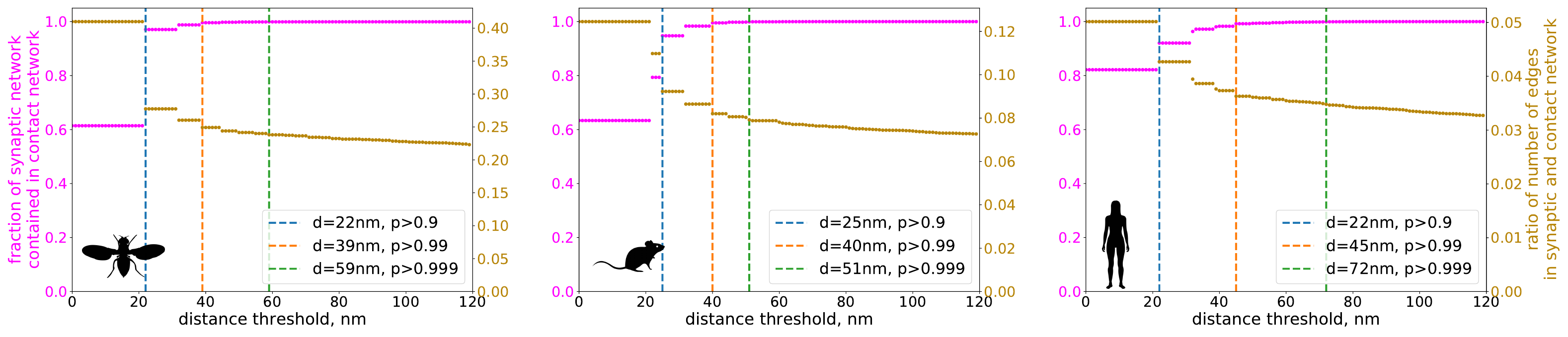}
    \caption{Thresholding the contact network. We demonstrate the fraction of the synaptic network that is contained in the contact network for different thresholds of distances corresponding to a contactome edge (magenta). We also show the ratios of the number of edges in the synaptic network to that of the contact network at different thresholds (dark yellow). Finally, we provide the distance values at which $90$, $99$, and $99.9\%$ of the synaptic network is contained in the contactome (vertical lines). To construct our final contact network, we find the union of the synaptic network and the contact network thresholded at the distance corresponding to $99\%$ of the synaptic edges contained in the contact network.}
    \label{fig: contact network construction}
\end{figure}

Establishing synapses between neurons relies on their surfaces being in close spatial proximity. This spatial constraint can be expressed by constructing physical contact networks---contactomes.
We obtain the contactomes using the vertices of neuron meshes at lod=1 for all three organisms. 
For each pair of neurons, we calculated the smallest pairwise distance $d_{ij,\text{min}}$. Then, we applied the distance threshold $d_{ij,\text{min}}<d_{\text{thr}}$ under which more than 99\% of the synaptic edges are contained in the contact network, see the magenta line in \cref{fig: contact network construction}.
Rounding up to the nearest nanometer, we get the thresholds of $d_{\text{thr,fly}}=40\text{nm}$, $d_{\text{thr,mouse}}=41\text{nm}$, and $d_{\text{thr,human}}=46\text{nm}$. All the edges representing pairs of neurons at a distance under this distance threshold are assigned to the contact network. Additionally, we add the 1\% of the synaptic edges that are not currently contained in the contact network.

As expected from our procedure, the data-driven distance thresholds we obtained are in the order of magnitude of the size of the synaptic clefts in chemical synapses \cite{zuber2005mammalian}. These thresholds are much larger than the size of the intercellular space in gap junctions \cite{maeda2009structure}. Thus, contactome edges contain more than one important mode of inter-neuronal communication.
On the other hand, the thresholds we impose are significantly smaller than micrometer-level thresholds in Ref.~\cite{reimann2015algorithm}---there, the distance corresponds to the linear dimension of the dendritic spines. Even at the strict threshold values we imposed, most of the contactome edges do not correspond to the presence of a chemical synapse, as demonstrated by the network densities in \cref{table: basic info}. The connectome edges comprise 25\%, 8\%, and 0.8\% of the contactome edges in fly, mouse, and human respectively.

The distance dependence in the contactomes is well approximated by an exponential. We estimate the distance scale $d_0$, assuming $p(d)\propto e^{-d/d_0}$, in the three organisms: $d_0=12$, $12$, and $15$ soma sizes for fly, mouse, and human respectively. Remarkably, these distance values are similar to those estimated for the connectome (\cref{subsec: spatial aspects} and \cref{table: orientation}).

The contact networks we constructed are undirected and unweighted, but their construction can be generalized to account for these important structural aspects. For instance, weights could be assigned based on the area of the physical contact or the number of disconnected cell sub-regions in contact with each other for each pair of cells.

\subsection{Degree preserving model with and without the contact constraint}\label{subsec: ME}

The entropy of a network ensemble is defined as 
\begin{align}\label{eq: ent}
S(G) = - \sum\limits_{G}P(G)\ln P(G),
\end{align}
where the network ensemble assigns a probability $P(G)$ to each network $G$.
Maximum entropy network ensembles are a useful tool in network science. Inspired by information theory and statistical physics, they represent our partial knowledge of the network by preserving a set of constraints---e.g., degree sequence---without making any assumptions about any other aspects of its structure.

As synaptic network data may be noisy, incomplete, and variable in time due to synaptic plasticity, we capture its structural properties by using the \textit{canonical ensemble} that preserves network constraints on average. For instance, let soft constraints be the degree sequence $k_1^*,...,k_N^*$ (model \textbf{k}, referred to as \textit{soft configuration model} in literature). Then, 
\begin{align}\label{eq: constr}
    \left< k_i\right>=\sum\limits_j A(G)_{ij}P(G)=k_i^*,
\end{align}
where $A(G)_{ij}$ stands for the elements of the adjacency matrix $A$ of a network $G$. Another constraint comes from the fact that probabilities $P(G)$ should add up to $1$. Maximizing the entropy in \cref{eq: ent} with these constraints using the method of Lagrange multipliers \cite{park2004statistical}, we arrive at
\begin{align}\label{eq: pg}
    P(G) = \dfrac{e^{-H(G)}}{Z}, 
\end{align}
where $H(G) \equiv \sum\limits_i{\theta_i}k_i(G)=
\sum\limits_{i<j}(\theta_i+\theta_j)A(G)_{ij}$ is the Hamiltonian, and $Z \equiv \sum\limits_G e^{-H(G)} = \prod\limits_{i<j}(1+e^{-\theta_i-\theta_j})$ is the partition function. From \cref{eq: constr}, the explicit expressions for the Lagrange multipliers for the maximum entropy model can be obtained from $k_i^*=-\dfrac{1}{Z}\dfrac{\partial Z}{\partial \theta_i}$. Explicitly,
\begin{align}
    k_{i}^*=\sum\limits_{j} p_{ij} = \sum\limits_j\dfrac{e^{-\theta_i-\theta_j}}{1+e^{-\theta_i-\theta_j}},
\end{align}
where $p_{ij}$ is the probability of $i$ and $j$ forming an edge. Note that these edge probabilities are independent of each other---thus, the network ensemble can be sampled by forming each edge $ij$ with probability $p_{ij}$.
The parameters $\theta_i$ can be calculated using a simple and efficient iterative scheme with an update rule
\begin{align}
x_i^{t+1}=\dfrac{1}{k_i^*}\sum\limits_j \dfrac{1}{1/x_i^t+x_j^t},    
\end{align}
where $x_i\equiv e^{\theta_i}$, for $k_i^*>0$ \cite{vallarano2021fast}. If $k_i^*=0$, $p_{ij}=0$ in this ensemble.
%When the iterative scheme fails, we can instead estimate the parameters by maximizing the log-likelihood $\mathcal L(G^*|\theta)=-H(G^*,\theta)-\ln Z(\theta)$.The partial derivatives of the log-likelihood function with respect to the parameters $\theta_i$ are $\dfrac{\partial\mathcal L(G^*|\theta)}{\partial \theta_i} = \dfrac{\partial\mathcal F}{\partial \theta_i}-C_i$, thus the function is maximized precisely at the correct values of the Lagrange multipliers.

Contact constraints can be explicitly incorporated into the maximum entropy framework as hard constraints (model \textbf{k+c}): no edges (synapses) can exist between the nodes (neurons) that are not in physical contact.
There, we explicitly take the hard contact constraint into account by forcing $p_{ij}$ to be zero where no edge exists in the contact network. Then, the partition function is $Z=\prod\limits_{i<j,i\sim j}(1+e^{-\theta_i+\theta_j})$, where $i\sim j$ indicates the existence of an edge in the contactome \cite{kovacs2020uncovering}. The Lagrange multipliers $\theta_i$ and edge probabilities $p_{ij}$ between the nodes in physical contact can then be obtained in the same way as for model \textbf{k}.

%Note that the other simple models we use can be considered maximum entropy models as well, e.g., the maximum entropy randomization preserving the total number of edges corresponds to assigning equal probabilities to the edges, and the maximum entropy randomization preserving the distance dependence can be obtained by simply assigning the edge probabilities according to that distance dependence.

\subsection{Degree and wiring length preserving randomization}\label{subsec: exponetial distance preserving}

In spatial networks such as the brain, forming and maintaining edges is often associated with a wiring cost \cite{bullmore2012economy}. Possibly the simplest cost function is the linear function $f(d_{ij}) = d_{ij}$, where $d_{ij}$ is the distance between the nodes. Following Ref.~\cite{halu2014emergence}, in our model \textbf{k+L} we impose soft constraints on the total wiring cost $L$ in addition to preserving the degree distribution (\cref{eq: constr}):
\begin{align}\label{eq: L}
    \left<L\right> = \sum\limits_{i<j}A(G)_{ij} d_{ij}P(G) = L^*.
\end{align}
Maximizing the entropy of the network ensemble (\cref{eq: ent}) with constraints (\cref{eq: L} and \cref{eq: constr}) using the method of Lagrange multipliers similarly to \cref{subsec: ME}, we arrive at
\begin{align}
    k_i^*=\sum\limits_j p_{ij} = \sum\limits_j\dfrac{e^{- d_{ij}/d_0}e^{-\theta_i-\theta_j}}{1+e^{- d_{ij}/d_0}e^{-\theta_i-\theta_j}}.
\end{align}
To identify the parameters $d_0$ and $\theta_i$ (via $x_i\equiv e^{\theta_i}$), we ran the iterative procedure
\begin{align}
x_i^{t+1}=\dfrac{1}{k_i}\sum\limits_j \dfrac{1}{1/x_i^t+x_j^t e^{d/d_0}},    
\end{align}
for a range of length scales $d_0$ and found the parameter that preserves the total edge length (error below $.01\%$ of the total edge length for all three organisms). The distance dependence is not a soft or hard constraint that is explicitly preserved in this case, but the distance dependence at the optimal $d_0$ value (purple line in \cref{fig: randomizations}) is similar to the one obtained from data (dark blue line in \cref{fig: randomizations}), especially for the mouse. This suggests that restricting the total wiring length might be an important mechanism in shaping the structure of the connectome and that $f(d_{ij})=d_{ij}$ is a meaningful representation of the wiring cost. 

Interestingly, the parameter $d_0$ that matches the total cost of edges is similar in the fly, mouse, and human ($d_0=$9, 9, and 10
%$d_0=$9.055, 8.716, and 9.985
soma size units, respectively). That confirms that soma sizes define a spatial scale relevant to synapse formation.

%\as{A natural refinement of the \textbf{k+L} model is a model that preserves the degree sequence \textbf{k} together with the distance distribution \textbf{d}. }

\subsection{Network measures}\label{subsec: network measures}

\begin{figure}
    \centering
    \includegraphics[width=1.\linewidth]{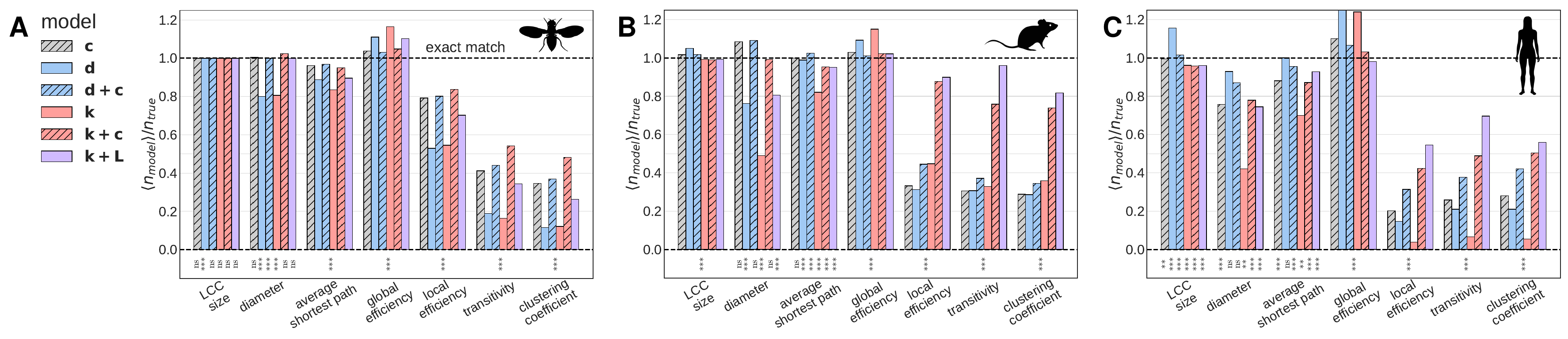}
    \caption{Comparing the basic network properties in empirical connectomes and models. For each model, the property values are averaged over 100 instances, except for the local clustering coefficient of the fly, which was averaged over 5 instances. The definitions of network properties are provided in \cref{subsec: network measures}.}
    \label{fig: fold changes extra}
\end{figure}

Here, we discuss the network measures we use to determine how well the models generalize beyond the explicitly built-in features \cite{dichio2024exploration}. We show the values of these measures for the connectome datasets in \cref{table: less basic info} and compare them to models by plotting the inverse fold change in \cref{fig: fold changes extra}. In this section, instead of the Euclidean distance between the nodes, we use the \textit{geodesic distance} $\delta_{ij}$, defined as the length of the paths between nodes $i$ and $j$ that contain the minimal number of edges. E.g., if node $i$ is not connected to node $j$, but they share a common neighbor $k$, $\delta_{ij}=2$. \\

We use the following measures:
\begin{itemize}
\itemsep-.3em 
    \item Size of the \textbf{largest connected component} ($\#$ nodes in LCC in \cref{table: less basic info})---the largest set of nodes connected by paths.
    \item Network \textbf{diameter} (diam. in \cref{table: less basic info})---the largest shortest distance between the nodes in the network $\max(\delta_{ij})$.
    \item  \textbf{Average shortest path} (average sh. path in \cref{table: less basic info}) refers to the average geodesic distance $\dfrac{1}{N}\sum\limits_{i\neq j} \delta_{ij}$.
    \item \textbf{Global efficiency}---a quantity related to the harmonic mean of the geodesic distance defined as $E_\text{g} = \dfrac{1}{N(N-1)}\sum\limits_{i\neq j} \dfrac{1}{\delta_{ij}}$.
    \item \textbf{Local efficiency}---an average of the global efficiencies of the subgraphs induced by the neighbors of each node, defined as $E_{\text{loc}}=\dfrac{1}{N}\sum\limits_i E_\text{g}(G_i)$.
\end{itemize}

Additionally, we use the following network measures related to triadic closure that compare the number of triangles $n_\Delta$ to the number of triplets $n_{\Lambda}$:
\begin{itemize}
\itemsep-.4em 
    \item \textbf{Transitivity}---the ratio of the number of triangles and the number of triplets for the entire network: $T=\dfrac{n_\Delta}{n_\Lambda}$.
    \item \textbf{Average clustering coefficient} (clust. coeff. in \cref{table: less basic info})---the average $C=\sum\limits_i C_i$ of the clustering coefficients $C_i$ of individual nodes. The individual clustering coefficients are defined as $C_i=\dfrac{n_{\Delta_i}}{n_{\Lambda_i}}$---the ratio of the number of triangles involving node $i$ to the number of pairs of its neighbors. If the node has less than 2 neighbors ($k_i<2$), its clustering coefficient is assigned to 0.
\end{itemize}

The measures we use are standard tools in network neuroscience \cite{rubinov2010complex} and beyond \cite{latora2017complex}. Their calculation is implemented in the \textsc{NetworkX Python} package \cite{hagberg2008exploring}.

\bibliographystyle{naturemag}
\bibliography{main}

\begin{thebibliography}{10}
\expandafter\ifx\csname url\endcsname\relax
  \def\url#1{\texttt{#1}}\fi
\expandafter\ifx\csname urlprefix\endcsname\relax\def\urlprefix{URL }\fi
\providecommand{\bibinfo}[2]{#2}
\providecommand{\eprint}[2][]{\url{#2}}

\bibitem{telesford2011brain}
\bibinfo{author}{Telesford, Q.~K.}, \bibinfo{author}{Simpson, S.~L.},
  \bibinfo{author}{Burdette, J.~H.}, \bibinfo{author}{Hayasaka, S.} \&
  \bibinfo{author}{Laurienti, P.~J.}
\newblock \bibinfo{title}{The brain as a complex system: using network science
  as a tool for understanding the brain}.
\newblock \emph{\bibinfo{journal}{Brain connectivity}}
  \textbf{\bibinfo{volume}{1}}, \bibinfo{pages}{295--308}
  (\bibinfo{year}{2011}).

\bibitem{barabasi2023neuroscience}
\bibinfo{author}{Barab{\'a}si, D.~L.} \emph{et~al.}
\newblock \bibinfo{title}{Neuroscience needs network science}.
\newblock \emph{\bibinfo{journal}{Journal of Neuroscience}}
  \textbf{\bibinfo{volume}{43}}, \bibinfo{pages}{5989--5995}
  (\bibinfo{year}{2023}).

\bibitem{bassett2017network}
\bibinfo{author}{Bassett, D.~S.} \& \bibinfo{author}{Sporns, O.}
\newblock \bibinfo{title}{Network neuroscience}.
\newblock \emph{\bibinfo{journal}{Nature neuroscience}}
  \textbf{\bibinfo{volume}{20}}, \bibinfo{pages}{353--364}
  (\bibinfo{year}{2017}).

\bibitem{scannell1995analysis}
\bibinfo{author}{Scannell, J.~W.}, \bibinfo{author}{Blakemore, C.} \&
  \bibinfo{author}{Young, M.~P.}
\newblock \bibinfo{title}{Analysis of connectivity in the cat cerebral cortex}.
\newblock \emph{\bibinfo{journal}{Journal of Neuroscience}}
  \textbf{\bibinfo{volume}{15}}, \bibinfo{pages}{1463--1483}
  (\bibinfo{year}{1995}).

\bibitem{kotter2004online}
\bibinfo{author}{K{\"o}tter, R.}
\newblock \bibinfo{title}{Online retrieval, processing, and visualization of
  primate connectivity data from the {C}o{C}o{M}ac database}.
\newblock \emph{\bibinfo{journal}{Neuroinformatics}}
  \textbf{\bibinfo{volume}{2}}, \bibinfo{pages}{127--144}
  (\bibinfo{year}{2004}).

\bibitem{lanciego2011half}
\bibinfo{author}{Lanciego, J.~L.} \& \bibinfo{author}{Wouterlood, F.~G.}
\newblock \bibinfo{title}{A half century of experimental neuroanatomical
  tracing}.
\newblock \emph{\bibinfo{journal}{Journal of chemical neuroanatomy}}
  \textbf{\bibinfo{volume}{42}}, \bibinfo{pages}{157--183}
  (\bibinfo{year}{2011}).

\bibitem{markov2014weighted}
\bibinfo{author}{Markov, N.~T.} \emph{et~al.}
\newblock \bibinfo{title}{A weighted and directed interareal connectivity
  matrix for macaque cerebral cortex}.
\newblock \emph{\bibinfo{journal}{Cerebral cortex}}
  \textbf{\bibinfo{volume}{24}}, \bibinfo{pages}{17--36}
  (\bibinfo{year}{2014}).

\bibitem{oh2014mesoscale}
\bibinfo{author}{Oh, S.~W.} \emph{et~al.}
\newblock \bibinfo{title}{A mesoscale connectome of the mouse brain}.
\newblock \emph{\bibinfo{journal}{Nature}} \textbf{\bibinfo{volume}{508}},
  \bibinfo{pages}{207--214} (\bibinfo{year}{2014}).

\bibitem{zingg2014neural}
\bibinfo{author}{Zingg, B.} \emph{et~al.}
\newblock \bibinfo{title}{Neural networks of the mouse neocortex}.
\newblock \emph{\bibinfo{journal}{Cell}} \textbf{\bibinfo{volume}{156}},
  \bibinfo{pages}{1096--1111} (\bibinfo{year}{2014}).

\bibitem{chiang2011three}
\bibinfo{author}{Chiang, A.-S.} \emph{et~al.}
\newblock \bibinfo{title}{Three-dimensional reconstruction of brain-wide wiring
  networks in drosophila at single-cell resolution}.
\newblock \emph{\bibinfo{journal}{Current biology}}
  \textbf{\bibinfo{volume}{21}}, \bibinfo{pages}{1--11} (\bibinfo{year}{2011}).

\bibitem{bullmore2009complex}
\bibinfo{author}{Bullmore, E.} \& \bibinfo{author}{Sporns, O.}
\newblock \bibinfo{title}{Complex brain networks: graph theoretical analysis of
  structural and functional systems}.
\newblock \emph{\bibinfo{journal}{Nature reviews neuroscience}}
  \textbf{\bibinfo{volume}{10}}, \bibinfo{pages}{186--198}
  (\bibinfo{year}{2009}).

\bibitem{betzel2019distance}
\bibinfo{author}{Betzel, R.~F.}, \bibinfo{author}{Griffa, A.},
  \bibinfo{author}{Hagmann, P.} \& \bibinfo{author}{Mi{\v{s}}i{\'c}, B.}
\newblock \bibinfo{title}{Distance-dependent consensus thresholds for
  generating group-representative structural brain networks}.
\newblock \emph{\bibinfo{journal}{Network neuroscience}}
  \textbf{\bibinfo{volume}{3}}, \bibinfo{pages}{475--496}
  (\bibinfo{year}{2019}).

\bibitem{horvat2016spatial}
\bibinfo{author}{Horv{\'a}t, S.} \emph{et~al.}
\newblock \bibinfo{title}{Spatial embedding and wiring cost constrain the
  functional layout of the cortical network of rodents and primates}.
\newblock \emph{\bibinfo{journal}{PLoS biology}} \textbf{\bibinfo{volume}{14}},
  \bibinfo{pages}{e1002512} (\bibinfo{year}{2016}).

\bibitem{bullmore2012economy}
\bibinfo{author}{Bullmore, E.} \& \bibinfo{author}{Sporns, O.}
\newblock \bibinfo{title}{The economy of brain network organization}.
\newblock \emph{\bibinfo{journal}{Nature reviews neuroscience}}
  \textbf{\bibinfo{volume}{13}}, \bibinfo{pages}{336--349}
  (\bibinfo{year}{2012}).

\bibitem{bassett2010efficient}
\bibinfo{author}{Bassett, D.~S.} \emph{et~al.}
\newblock \bibinfo{title}{Efficient physical embedding of topologically complex
  information processing networks in brains and computer circuits}.
\newblock \emph{\bibinfo{journal}{PLoS computational biology}}
  \textbf{\bibinfo{volume}{6}}, \bibinfo{pages}{e1000748}
  (\bibinfo{year}{2010}).

\bibitem{chen2006wiring}
\bibinfo{author}{Chen, B.~L.}, \bibinfo{author}{Hall, D.~H.} \&
  \bibinfo{author}{Chklovskii, D.~B.}
\newblock \bibinfo{title}{Wiring optimization can relate neuronal structure and
  function}.
\newblock \emph{\bibinfo{journal}{Proceedings of the National Academy of
  Sciences}} \textbf{\bibinfo{volume}{103}}, \bibinfo{pages}{4723--4728}
  (\bibinfo{year}{2006}).

\bibitem{maertens2021multilayer}
\bibinfo{author}{Maertens, T.}, \bibinfo{author}{Sch{\"o}ll, E.},
  \bibinfo{author}{Ruiz, J.} \& \bibinfo{author}{H{\"o}vel, P.}
\newblock \bibinfo{title}{Multilayer network analysis of \emph{C. elegans}:
  {L}ooking into the locomotory circuitry}.
\newblock \emph{\bibinfo{journal}{Neurocomputing}}
  \textbf{\bibinfo{volume}{427}}, \bibinfo{pages}{238--261}
  (\bibinfo{year}{2021}).

\bibitem{yan2017network}
\bibinfo{author}{Yan, G.} \emph{et~al.}
\newblock \bibinfo{title}{Network control principles predict neuron function in
  the \emph{Caenorhabditis elegans} connectome}.
\newblock \emph{\bibinfo{journal}{Nature}} \textbf{\bibinfo{volume}{550}},
  \bibinfo{pages}{519--523} (\bibinfo{year}{2017}).

\bibitem{gushchin2015total}
\bibinfo{author}{Gushchin, A.} \& \bibinfo{author}{Tang, A.}
\newblock \bibinfo{title}{Total wiring length minimization of \emph{C. elegans}
  neural network: a constrained optimization approach}.
\newblock \emph{\bibinfo{journal}{PloS one}} \textbf{\bibinfo{volume}{10}},
  \bibinfo{pages}{e0145029} (\bibinfo{year}{2015}).

\bibitem{song2021maximum}
\bibinfo{author}{Song, Y.}, \bibinfo{author}{Zhou, D.} \& \bibinfo{author}{Li,
  S.}
\newblock \bibinfo{title}{Maximum entropy principle underlies wiring length
  distribution in brain networks}.
\newblock \emph{\bibinfo{journal}{Cerebral cortex}}
  \textbf{\bibinfo{volume}{31}}, \bibinfo{pages}{4628--4641}
  (\bibinfo{year}{2021}).

\bibitem{lynn2024heavy}
\bibinfo{author}{Lynn, C.~W.}, \bibinfo{author}{Holmes, C.~M.} \&
  \bibinfo{author}{Palmer, S.~E.}
\newblock \bibinfo{title}{Heavy-tailed neuronal connectivity arises from
  {H}ebbian self-organization}.
\newblock \emph{\bibinfo{journal}{Nature Physics}} \bibinfo{pages}{1--8}
  (\bibinfo{year}{2024}).

\bibitem{haber2023structure}
\bibinfo{author}{Haber, A.}, \bibinfo{author}{Wanner, A.},
  \bibinfo{author}{Friedrich, R.~W.} \& \bibinfo{author}{Schneidman, E.}
\newblock \bibinfo{title}{The structure and function of neural connectomes are
  shaped by a small number of design principles}.
\newblock \emph{\bibinfo{journal}{bioRxiv}} \bibinfo{pages}{2023--03}
  (\bibinfo{year}{2023}).

\bibitem{helmstaedter2013cellular}
\bibinfo{author}{Helmstaedter, M.}
\newblock \bibinfo{title}{Cellular-resolution connectomics: challenges of dense
  neural circuit reconstruction}.
\newblock \emph{\bibinfo{journal}{Nature methods}}
  \textbf{\bibinfo{volume}{10}}, \bibinfo{pages}{501--507}
  (\bibinfo{year}{2013}).

\bibitem{hildebrand2017whole}
\bibinfo{author}{Hildebrand, D. G.~C.} \emph{et~al.}
\newblock \bibinfo{title}{Whole-brain serial-section electron microscopy in
  larval zebrafish}.
\newblock \emph{\bibinfo{journal}{Nature}} \textbf{\bibinfo{volume}{545}},
  \bibinfo{pages}{345--349} (\bibinfo{year}{2017}).

\bibitem{motta2019dense}
\bibinfo{author}{Motta, A.} \emph{et~al.}
\newblock \bibinfo{title}{Dense connectomic reconstruction in layer 4 of the
  somatosensory cortex}.
\newblock \emph{\bibinfo{journal}{Science}} \textbf{\bibinfo{volume}{366}},
  \bibinfo{pages}{eaay3134} (\bibinfo{year}{2019}).

\bibitem{winding2023connectome}
\bibinfo{author}{Winding, M.} \emph{et~al.}
\newblock \bibinfo{title}{The connectome of an insect brain}.
\newblock \emph{\bibinfo{journal}{Science}} \textbf{\bibinfo{volume}{379}},
  \bibinfo{pages}{eadd9330} (\bibinfo{year}{2023}).

\bibitem{dorkenwald2023neuronal}
\bibinfo{author}{Dorkenwald, S.} \emph{et~al.}
\newblock \bibinfo{title}{Neuronal wiring diagram of an adult brain}.
\newblock \emph{\bibinfo{journal}{bioRxiv}}  (\bibinfo{year}{2023}).

\bibitem{xu2020connectome}
\bibinfo{author}{Scheffer, L.~K.} \emph{et~al.}
\newblock \bibinfo{title}{A connectome and analysis of the adult
  \emph{Drosophila} central brain}.
\newblock \emph{\bibinfo{journal}{Elife}} \textbf{\bibinfo{volume}{9}},
  \bibinfo{pages}{e57443} (\bibinfo{year}{2020}).

\bibitem{bae2021functional}
\bibinfo{author}{Bae, J.~A.} \emph{et~al.}
\newblock \bibinfo{title}{Functional connectomics spanning multiple areas of
  mouse visual cortex}.
\newblock \emph{\bibinfo{journal}{bioRxiv}}  (\bibinfo{year}{2021}).

\bibitem{shapson2021connectomic}
\bibinfo{author}{Shapson-Coe, A.} \emph{et~al.}
\newblock \bibinfo{title}{A connectomic study of a petascale fragment of human
  cerebral cortex}.
\newblock \emph{\bibinfo{journal}{bioRxiv}}  (\bibinfo{year}{2021}).

\bibitem{van2016comparative}
\bibinfo{author}{Van~den Heuvel, M.~P.}, \bibinfo{author}{Bullmore, E.~T.} \&
  \bibinfo{author}{Sporns, O.}
\newblock \bibinfo{title}{Comparative connectomics}.
\newblock \emph{\bibinfo{journal}{Trends in cognitive sciences}}
  \textbf{\bibinfo{volume}{20}}, \bibinfo{pages}{345--361}
  (\bibinfo{year}{2016}).

\bibitem{azevedo2009equal}
\bibinfo{author}{Azevedo, F.~A.} \emph{et~al.}
\newblock \bibinfo{title}{Equal numbers of neuronal and nonneuronal cells make
  the human brain an isometrically scaled-up primate brain}.
\newblock \emph{\bibinfo{journal}{Journal of Comparative Neurology}}
  \textbf{\bibinfo{volume}{513}}, \bibinfo{pages}{532--541}
  (\bibinfo{year}{2009}).

\bibitem{herculano2006cellular}
\bibinfo{author}{Herculano-Houzel, S.}, \bibinfo{author}{Mota, B.} \&
  \bibinfo{author}{Lent, R.}
\newblock \bibinfo{title}{Cellular scaling rules for rodent brains}.
\newblock \emph{\bibinfo{journal}{Proceedings of the National Academy of
  Sciences}} \textbf{\bibinfo{volume}{103}}, \bibinfo{pages}{12138--12143}
  (\bibinfo{year}{2006}).

\bibitem{barthelemy2011spatial}
\bibinfo{author}{Barth{\'e}lemy, M.}
\newblock \bibinfo{title}{Spatial networks}.
\newblock \emph{\bibinfo{journal}{Physics reports}}
  \textbf{\bibinfo{volume}{499}}, \bibinfo{pages}{1--101}
  (\bibinfo{year}{2011}).

\bibitem{bassett2018spatial}
\bibinfo{author}{Bassett, D.~S.} \& \bibinfo{author}{Stiso, J.}
\newblock \bibinfo{title}{Spatial brain networks}.
\newblock \emph{\bibinfo{journal}{Comptes Rendus Physique}}
  \textbf{\bibinfo{volume}{19}}, \bibinfo{pages}{253--264}
  (\bibinfo{year}{2018}).

\bibitem{markov2011weight}
\bibinfo{author}{Markov, N.~T.} \emph{et~al.}
\newblock \bibinfo{title}{Weight consistency specifies regularities of macaque
  cortical networks}.
\newblock \emph{\bibinfo{journal}{Cerebral cortex}}
  \textbf{\bibinfo{volume}{21}}, \bibinfo{pages}{1254--1272}
  (\bibinfo{year}{2011}).

\bibitem{markov2013role}
\bibinfo{author}{Markov, N.~T.} \emph{et~al.}
\newblock \bibinfo{title}{The role of long-range connections on the specificity
  of the macaque interareal cortical network}.
\newblock \emph{\bibinfo{journal}{Proceedings of the National Academy of
  Sciences}} \textbf{\bibinfo{volume}{110}}, \bibinfo{pages}{5187--5192}
  (\bibinfo{year}{2013}).

\bibitem{smith2021neurons}
\bibinfo{author}{Smith, J.~H.} \emph{et~al.}
\newblock \bibinfo{title}{How neurons exploit fractal geometry to optimize
  their network connectivity}.
\newblock \emph{\bibinfo{journal}{Scientific reports}}
  \textbf{\bibinfo{volume}{11}}, \bibinfo{pages}{2332} (\bibinfo{year}{2021}).

\bibitem{ansell2023unveiling}
\bibinfo{author}{Ansell, H.~S.} \& \bibinfo{author}{Kov{\'a}cs, I.~A.}
\newblock \bibinfo{title}{Unveiling universal aspects of the cellular anatomy
  of the brain}.
\newblock \emph{\bibinfo{journal}{arXiv preprint arXiv:2306.12289}}
  (\bibinfo{year}{2023}).

\bibitem{pete2023network}
\bibinfo{author}{Pete, G.}, \bibinfo{author}{Tim{\'a}r, {\'A}.},
  \bibinfo{author}{Stef{\'a}nsson, S.~{\"O}.}, \bibinfo{author}{Bonamassa, I.}
  \& \bibinfo{author}{P{\'o}sfai, M.}
\newblock \bibinfo{title}{A network-of-networks model for physical networks}.
\newblock \emph{\bibinfo{journal}{arXiv preprint arXiv:2306.01583}}
  (\bibinfo{year}{2023}).

\bibitem{posfai2023impact}
\bibinfo{author}{P{\'o}sfai, M.} \emph{et~al.}
\newblock \bibinfo{title}{Impact of physicality on network structure}.
\newblock \emph{\bibinfo{journal}{Nature Physics}} \bibinfo{pages}{1--8}
  (\bibinfo{year}{2023}).

\bibitem{amaral2000classes}
\bibinfo{author}{Amaral, L. A.~N.}, \bibinfo{author}{Scala, A.},
  \bibinfo{author}{Barthelemy, M.} \& \bibinfo{author}{Stanley, H.~E.}
\newblock \bibinfo{title}{Classes of small-world networks}.
\newblock \emph{\bibinfo{journal}{Proceedings of the national academy of
  sciences}} \textbf{\bibinfo{volume}{97}}, \bibinfo{pages}{11149--11152}
  (\bibinfo{year}{2000}).

\bibitem{barabasi1999emergence}
\bibinfo{author}{Barab{\'a}si, A.-L.} \& \bibinfo{author}{Albert, R.}
\newblock \bibinfo{title}{Emergence of scaling in random networks}.
\newblock \emph{\bibinfo{journal}{science}} \textbf{\bibinfo{volume}{286}},
  \bibinfo{pages}{509--512} (\bibinfo{year}{1999}).

\bibitem{lynn2022emergent}
\bibinfo{author}{Lynn, C.~W.}, \bibinfo{author}{Holmes, C.~M.} \&
  \bibinfo{author}{Palmer, S.~E.}
\newblock \bibinfo{title}{Emergent scale-free networks}.
\newblock \emph{\bibinfo{journal}{arXiv preprint arXiv:2210.06453}}
  (\bibinfo{year}{2022}).

\bibitem{peters1976projection}
\bibinfo{author}{Peters, A.} \& \bibinfo{author}{Feldman, M.~L.}
\newblock \bibinfo{title}{The projection of the lateral geniculate nucleus to
  area 17 of the rat cerebral cortex. {I}. {G}eneral description}.
\newblock \emph{\bibinfo{journal}{Journal of neurocytology}}
  \textbf{\bibinfo{volume}{5}}, \bibinfo{pages}{63--84} (\bibinfo{year}{1976}).

\bibitem{braitenberg1991peters}
\bibinfo{author}{Braitenberg, V.}, \bibinfo{author}{Sch{\"u}z, A.},
  \bibinfo{author}{Braitenberg, V.} \& \bibinfo{author}{Sch{\"u}z, A.}
\newblock \bibinfo{title}{Peters’ rule and {W}hite’s exceptions}.
\newblock \emph{\bibinfo{journal}{Anatomy of the Cortex: Statistics and
  Geometry}} \bibinfo{pages}{109--112} (\bibinfo{year}{1991}).

\bibitem{rees2017weighing}
\bibinfo{author}{Rees, C.~L.}, \bibinfo{author}{Moradi, K.} \&
  \bibinfo{author}{Ascoli, G.~A.}
\newblock \bibinfo{title}{Weighing the evidence in {P}eters’ rule: does
  neuronal morphology predict connectivity?}
\newblock \emph{\bibinfo{journal}{Trends in neurosciences}}
  \textbf{\bibinfo{volume}{40}}, \bibinfo{pages}{63--71}
  (\bibinfo{year}{2017}).

\bibitem{stepanyants2005neurogeometry}
\bibinfo{author}{Stepanyants, A.} \& \bibinfo{author}{Chklovskii, D.~B.}
\newblock \bibinfo{title}{Neurogeometry and potential synaptic connectivity}.
\newblock \emph{\bibinfo{journal}{Trends in neurosciences}}
  \textbf{\bibinfo{volume}{28}}, \bibinfo{pages}{387--394}
  (\bibinfo{year}{2005}).

\bibitem{reimann2015algorithm}
\bibinfo{author}{Reimann, M.~W.}, \bibinfo{author}{King, J.~G.},
  \bibinfo{author}{Muller, E.~B.}, \bibinfo{author}{Ramaswamy, S.} \&
  \bibinfo{author}{Markram, H.}
\newblock \bibinfo{title}{An algorithm to predict the connectome of neural
  microcircuits}.
\newblock \emph{\bibinfo{journal}{Frontiers in computational neuroscience}}
  \textbf{\bibinfo{volume}{9}}, \bibinfo{pages}{120} (\bibinfo{year}{2015}).

\bibitem{kovacs2020uncovering}
\bibinfo{author}{Kov{\'a}cs, I.~A.}, \bibinfo{author}{Barab{\'a}si, D.~L.} \&
  \bibinfo{author}{Barab{\'a}si, A.-L.}
\newblock \bibinfo{title}{Uncovering the genetic blueprint of the \emph{C.
  elegans} nervous system}.
\newblock \emph{\bibinfo{journal}{Proceedings of the National Academy of
  Sciences}} \textbf{\bibinfo{volume}{117}}, \bibinfo{pages}{33570--33577}
  (\bibinfo{year}{2020}).

\bibitem{arnatkevicute2018hub}
\bibinfo{author}{Arnatkevi{\v{c}}i{\=u}t{\.e}, A.}, \bibinfo{author}{Fulcher,
  B.~D.}, \bibinfo{author}{Pocock, R.} \& \bibinfo{author}{Fornito, A.}
\newblock \bibinfo{title}{Hub connectivity, neuronal diversity, and gene
  expression in the \emph{Caenorhabditis elegans} connectome}.
\newblock \emph{\bibinfo{journal}{PLoS computational biology}}
  \textbf{\bibinfo{volume}{14}}, \bibinfo{pages}{e1005989}
  (\bibinfo{year}{2018}).

\bibitem{yin2020network}
\bibinfo{author}{Yin, C.} \emph{et~al.}
\newblock \bibinfo{title}{Network science characteristics of brain-derived
  neuronal cultures deciphered from quantitative phase imaging data}.
\newblock \emph{\bibinfo{journal}{Scientific reports}}
  \textbf{\bibinfo{volume}{10}}, \bibinfo{pages}{1--13} (\bibinfo{year}{2020}).

\bibitem{fulcher2016transcriptional}
\bibinfo{author}{Fulcher, B.~D.} \& \bibinfo{author}{Fornito, A.}
\newblock \bibinfo{title}{A transcriptional signature of hub connectivity in
  the mouse connectome}.
\newblock \emph{\bibinfo{journal}{Proceedings of the National Academy of
  Sciences}} \textbf{\bibinfo{volume}{113}}, \bibinfo{pages}{1435--1440}
  (\bibinfo{year}{2016}).

\bibitem{fornito2019bridging}
\bibinfo{author}{Fornito, A.}, \bibinfo{author}{Arnatkevi{\v{c}}i{\=u}t{\.e},
  A.} \& \bibinfo{author}{Fulcher, B.~D.}
\newblock \bibinfo{title}{Bridging the gap between connectome and
  transcriptome}.
\newblock \emph{\bibinfo{journal}{Trends in cognitive sciences}}
  \textbf{\bibinfo{volume}{23}}, \bibinfo{pages}{34--50}
  (\bibinfo{year}{2019}).

\bibitem{park2004statistical}
\bibinfo{author}{Park, J.} \& \bibinfo{author}{Newman, M.~E.}
\newblock \bibinfo{title}{Statistical mechanics of networks}.
\newblock \emph{\bibinfo{journal}{Physical Review E}}
  \textbf{\bibinfo{volume}{70}}, \bibinfo{pages}{066117}
  (\bibinfo{year}{2004}).

\bibitem{bianconi2021information}
\bibinfo{author}{Bianconi, G.}
\newblock \bibinfo{title}{Information theory of spatial network ensembles}.
\newblock In \emph{\bibinfo{booktitle}{Handbook on Entropy, Complexity and
  Spatial Dynamics}}, \bibinfo{pages}{61--96} (\bibinfo{publisher}{Edward Elgar
  Publishing}, \bibinfo{year}{2021}).

\bibitem{robins2007introduction}
\bibinfo{author}{Robins, G.}, \bibinfo{author}{Pattison, P.},
  \bibinfo{author}{Kalish, Y.} \& \bibinfo{author}{Lusher, D.}
\newblock \bibinfo{title}{An introduction to exponential random graph (p*)
  models for social networks}.
\newblock \emph{\bibinfo{journal}{Social networks}}
  \textbf{\bibinfo{volume}{29}}, \bibinfo{pages}{173--191}
  (\bibinfo{year}{2007}).

\bibitem{dichio2023statistical}
\bibinfo{author}{Dichio, V.} \& \bibinfo{author}{Fallani, F. D.~V.}
\newblock \bibinfo{title}{Statistical models of complex brain networks: a
  maximum entropy approach}.
\newblock \emph{\bibinfo{journal}{Reports on progress in physics}}
  (\bibinfo{year}{2023}).

\bibitem{witvliet2021connectomes}
\bibinfo{author}{Witvliet, D.} \emph{et~al.}
\newblock \bibinfo{title}{Connectomes across development reveal principles of
  brain maturation}.
\newblock \emph{\bibinfo{journal}{Nature}} \textbf{\bibinfo{volume}{596}},
  \bibinfo{pages}{257--261} (\bibinfo{year}{2021}).

\bibitem{hiesinger2018evolution}
\bibinfo{author}{Hiesinger, P.~R.} \& \bibinfo{author}{Hassan, B.~A.}
\newblock \bibinfo{title}{The evolution of variability and robustness in neural
  development}.
\newblock \emph{\bibinfo{journal}{Trends in Neurosciences}}
  \textbf{\bibinfo{volume}{41}}, \bibinfo{pages}{577--586}
  (\bibinfo{year}{2018}).

\bibitem{schlegel2023whole}
\bibinfo{author}{Schlegel, P.} \emph{et~al.}
\newblock \bibinfo{title}{Whole-brain annotation and multi-connectome cell
  typing quantifies circuit stereotypy in \emph{Drosophila}}.
\newblock \emph{\bibinfo{journal}{bioRxiv}} \bibinfo{pages}{2023--06}
  (\bibinfo{year}{2023}).

\bibitem{hao2023proper}
\bibinfo{author}{Hao, B.} \& \bibinfo{author}{Kov{\'a}cs, I.~A.}
\newblock \bibinfo{title}{Proper network randomization is key to assessing
  social balance}.
\newblock \emph{\bibinfo{journal}{arXiv preprint arXiv:2305.16561}}
  (\bibinfo{year}{2023}).

\bibitem{kivela2014multilayer}
\bibinfo{author}{Kivel{\"a}, M.} \emph{et~al.}
\newblock \bibinfo{title}{Multilayer networks}.
\newblock \emph{\bibinfo{journal}{Journal of complex networks}}
  \textbf{\bibinfo{volume}{2}}, \bibinfo{pages}{203--271}
  (\bibinfo{year}{2014}).

\bibitem{ahn2006wiring}
\bibinfo{author}{Ahn, Y.-Y.}, \bibinfo{author}{Jeong, H.} \&
  \bibinfo{author}{Kim, B.~J.}
\newblock \bibinfo{title}{Wiring cost in the organization of a biological
  neuronal network}.
\newblock \emph{\bibinfo{journal}{Physica A: Statistical Mechanics and its
  Applications}} \textbf{\bibinfo{volume}{367}}, \bibinfo{pages}{531--537}
  (\bibinfo{year}{2006}).

\bibitem{halu2014emergence}
\bibinfo{author}{Halu, A.}, \bibinfo{author}{Mukherjee, S.} \&
  \bibinfo{author}{Bianconi, G.}
\newblock \bibinfo{title}{Emergence of overlap in ensembles of spatial
  multiplexes and statistical mechanics of spatial interacting network
  ensembles}.
\newblock \emph{\bibinfo{journal}{Physical Review E}}
  \textbf{\bibinfo{volume}{89}}, \bibinfo{pages}{012806}
  (\bibinfo{year}{2014}).

\bibitem{milo2002network}
\bibinfo{author}{Milo, R.} \emph{et~al.}
\newblock \bibinfo{title}{Network motifs: simple building blocks of complex
  networks}.
\newblock \emph{\bibinfo{journal}{Science}} \textbf{\bibinfo{volume}{298}},
  \bibinfo{pages}{824--827} (\bibinfo{year}{2002}).

\bibitem{prvzulj2007biological}
\bibinfo{author}{Pr{\v{z}}ulj, N.}
\newblock \bibinfo{title}{Biological network comparison using graphlet degree
  distribution}.
\newblock \emph{\bibinfo{journal}{Bioinformatics}}
  \textbf{\bibinfo{volume}{23}}, \bibinfo{pages}{e177--e183}
  (\bibinfo{year}{2007}).

\bibitem{kovacs2019network}
\bibinfo{author}{Kov{\'a}cs, I.~A.} \emph{et~al.}
\newblock \bibinfo{title}{Network-based prediction of protein interactions}.
\newblock \emph{\bibinfo{journal}{Nature communications}}
  \textbf{\bibinfo{volume}{10}}, \bibinfo{pages}{1240} (\bibinfo{year}{2019}).

\bibitem{rubinov2010complex}
\bibinfo{author}{Rubinov, M.} \& \bibinfo{author}{Sporns, O.}
\newblock \bibinfo{title}{Complex network measures of brain connectivity: uses
  and interpretations}.
\newblock \emph{\bibinfo{journal}{Neuroimage}} \textbf{\bibinfo{volume}{52}},
  \bibinfo{pages}{1059--1069} (\bibinfo{year}{2010}).

\bibitem{latora2001efficient}
\bibinfo{author}{Latora, V.} \& \bibinfo{author}{Marchiori, M.}
\newblock \bibinfo{title}{Efficient behavior of small-world networks}.
\newblock \emph{\bibinfo{journal}{Physical review letters}}
  \textbf{\bibinfo{volume}{87}}, \bibinfo{pages}{198701}
  (\bibinfo{year}{2001}).

\bibitem{maeda2009structure}
\bibinfo{author}{Maeda, S.} \emph{et~al.}
\newblock \bibinfo{title}{Structure of the connexin 26 gap junction channel at
  3.5 {\aa} resolution}.
\newblock \emph{\bibinfo{journal}{Nature}} \textbf{\bibinfo{volume}{458}},
  \bibinfo{pages}{597--602} (\bibinfo{year}{2009}).

\bibitem{udvary2022impact}
\bibinfo{author}{Udvary, D.} \emph{et~al.}
\newblock \bibinfo{title}{The impact of neuron morphology on cortical network
  architecture}.
\newblock \emph{\bibinfo{journal}{Cell Reports}} \textbf{\bibinfo{volume}{39}}
  (\bibinfo{year}{2022}).

\bibitem{oldham2022modeling}
\bibinfo{author}{Oldham, S.} \emph{et~al.}
\newblock \bibinfo{title}{Modeling spatial, developmental, physiological, and
  topological constraints on human brain connectivity}.
\newblock \emph{\bibinfo{journal}{Science advances}}
  \textbf{\bibinfo{volume}{8}}, \bibinfo{pages}{eabm6127}
  (\bibinfo{year}{2022}).

\bibitem{kaiser2004modelling}
\bibinfo{author}{Kaiser, M.} \& \bibinfo{author}{Hilgetag, C.~C.}
\newblock \bibinfo{title}{Modelling the development of cortical systems
  networks}.
\newblock \emph{\bibinfo{journal}{Neurocomputing}}
  \textbf{\bibinfo{volume}{58}}, \bibinfo{pages}{297--302}
  (\bibinfo{year}{2004}).

\bibitem{vallarano2021fast}
\bibinfo{author}{Vallarano, N.} \emph{et~al.}
\newblock \bibinfo{title}{Fast and scalable likelihood maximization for
  exponential random graph models with local constraints}.
\newblock \emph{\bibinfo{journal}{Scientific Reports}}
  \textbf{\bibinfo{volume}{11}}, \bibinfo{pages}{1--33} (\bibinfo{year}{2021}).

\bibitem{abbott2020mind}
\bibinfo{author}{Abbott, L.~F.} \emph{et~al.}
\newblock \bibinfo{title}{The mind of a mouse}.
\newblock \emph{\bibinfo{journal}{Cell}} \textbf{\bibinfo{volume}{182}},
  \bibinfo{pages}{1372--1376} (\bibinfo{year}{2020}).

\bibitem{bianconi2009entropy}
\bibinfo{author}{Bianconi, G.}
\newblock \bibinfo{title}{Entropy of network ensembles}.
\newblock \emph{\bibinfo{journal}{Physical Review E}}
  \textbf{\bibinfo{volume}{79}}, \bibinfo{pages}{036114}
  (\bibinfo{year}{2009}).

\bibitem{gallo2023strong}
\bibinfo{author}{Gallo, A.}, \bibinfo{author}{Garlaschelli, D.},
  \bibinfo{author}{Lambiotte, R.}, \bibinfo{author}{Saracco, F.} \&
  \bibinfo{author}{Squartini, T.}
\newblock \bibinfo{title}{Strong, weak or no balance? {T}esting structural
  hypotheses against real networks}.
\newblock \emph{\bibinfo{journal}{arXiv preprint arXiv:2303.07023}}
  (\bibinfo{year}{2023}).

\bibitem{saracco2022entropy}
\bibinfo{author}{Saracco, F.}, \bibinfo{author}{Petri, G.},
  \bibinfo{author}{Lambiotte, R.} \& \bibinfo{author}{Squartini, T.}
\newblock \bibinfo{title}{Entropy-based random models for hypergraphs}.
\newblock \emph{\bibinfo{journal}{arXiv preprint arXiv:2207.12123}}
  (\bibinfo{year}{2022}).

\bibitem{lizier2023analytic}
\bibinfo{author}{Lizier, J.~T.}, \bibinfo{author}{Bauer, F.},
  \bibinfo{author}{Atay, F.~M.} \& \bibinfo{author}{Jost, J.}
\newblock \bibinfo{title}{Analytic relationship of relative synchronizability
  to network structure and motifs}.
\newblock \emph{\bibinfo{journal}{arXiv preprint arXiv:2305.10509}}
  (\bibinfo{year}{2023}).

\bibitem{kim2012spatiotemporal}
\bibinfo{author}{Kim, M.-S.}, \bibinfo{author}{Kim, J.-R.},
  \bibinfo{author}{Kim, D.}, \bibinfo{author}{Lander, A.~D.} \&
  \bibinfo{author}{Cho, K.-H.}
\newblock \bibinfo{title}{Spatiotemporal network motif reveals the biological
  traits of developmental gene regulatory networks in drosophila melanogaster}.
\newblock \emph{\bibinfo{journal}{BMC systems biology}}
  \textbf{\bibinfo{volume}{6}}, \bibinfo{pages}{1--10} (\bibinfo{year}{2012}).

\bibitem{fields2015glial}
\bibinfo{author}{Fields, R.~D.}, \bibinfo{author}{Woo, D.~H.} \&
  \bibinfo{author}{Basser, P.~J.}
\newblock \bibinfo{title}{Glial regulation of the neuronal connectome through
  local and long-distant communication}.
\newblock \emph{\bibinfo{journal}{Neuron}} \textbf{\bibinfo{volume}{86}},
  \bibinfo{pages}{374--386} (\bibinfo{year}{2015}).

\bibitem{schneider2023cell}
\bibinfo{author}{Schneider-Mizell, C.~M.} \emph{et~al.}
\newblock \bibinfo{title}{Cell-type-specific inhibitory circuitry from a
  connectomic census of mouse visual cortex}.
\newblock \emph{\bibinfo{journal}{bioRxiv}}  (\bibinfo{year}{2023}).

\bibitem{elabbady2022quantitative}
\bibinfo{author}{Elabbady, L.} \emph{et~al.}
\newblock \bibinfo{title}{Quantitative census of local somatic features in
  mouse visual cortex}.
\newblock \emph{\bibinfo{journal}{bioRxiv}} \bibinfo{pages}{2022--07}
  (\bibinfo{year}{2022}).

\bibitem{scikit-learn}
\bibinfo{author}{Pedregosa, F.} \emph{et~al.}
\newblock \bibinfo{title}{Scikit-learn: {M}achine learning in {P}ython}.
\newblock \emph{\bibinfo{journal}{Journal of Machine Learning Research}}
  \textbf{\bibinfo{volume}{12}}, \bibinfo{pages}{2825--2830}
  (\bibinfo{year}{2011}).

\bibitem{zuber2005mammalian}
\bibinfo{author}{Zuber, B.}, \bibinfo{author}{Nikonenko, I.},
  \bibinfo{author}{Klauser, P.}, \bibinfo{author}{Muller, D.} \&
  \bibinfo{author}{Dubochet, J.}
\newblock \bibinfo{title}{The mammalian central nervous synaptic cleft contains
  a high density of periodically organized complexes}.
\newblock \emph{\bibinfo{journal}{Proceedings of the National Academy of
  Sciences}} \textbf{\bibinfo{volume}{102}}, \bibinfo{pages}{19192--19197}
  (\bibinfo{year}{2005}).

\bibitem{dichio2024exploration}
\bibinfo{author}{Dichio, V.} \& \bibinfo{author}{Fallani, F. D.~V.}
\newblock \bibinfo{title}{Exploration-exploitation paradigm for networked
  biological systems}.
\newblock \emph{\bibinfo{journal}{Physical Review Letters}}
  \textbf{\bibinfo{volume}{132}}, \bibinfo{pages}{098402}
  (\bibinfo{year}{2024}).

\bibitem{latora2017complex}
\bibinfo{author}{Latora, V.}, \bibinfo{author}{Nicosia, V.} \&
  \bibinfo{author}{Russo, G.}
\newblock \emph{\bibinfo{title}{Complex networks: principles, methods and
  applications}} (\bibinfo{publisher}{Cambridge University Press},
  \bibinfo{year}{2017}).

\bibitem{hagberg2008exploring}
\bibinfo{author}{Hagberg, A.}, \bibinfo{author}{Swart, P.} \&
  \bibinfo{author}{S~Chult, D.}
\newblock \bibinfo{title}{Exploring network structure, dynamics, and function
  using {N}etwork{X}}.
\newblock \bibinfo{type}{Tech. Rep.}, \bibinfo{institution}{Los Alamos National
  Lab.(LANL), Los Alamos, NM (United States)} (\bibinfo{year}{2008}).

\end{thebibliography}

%\FloatBarrier
%\newpage
\section{Supplementary Information}
\beginsupplement
\section{Supplementary figures and tables}

\begin{figure}
    \centering
    \includegraphics[width=.8\linewidth]{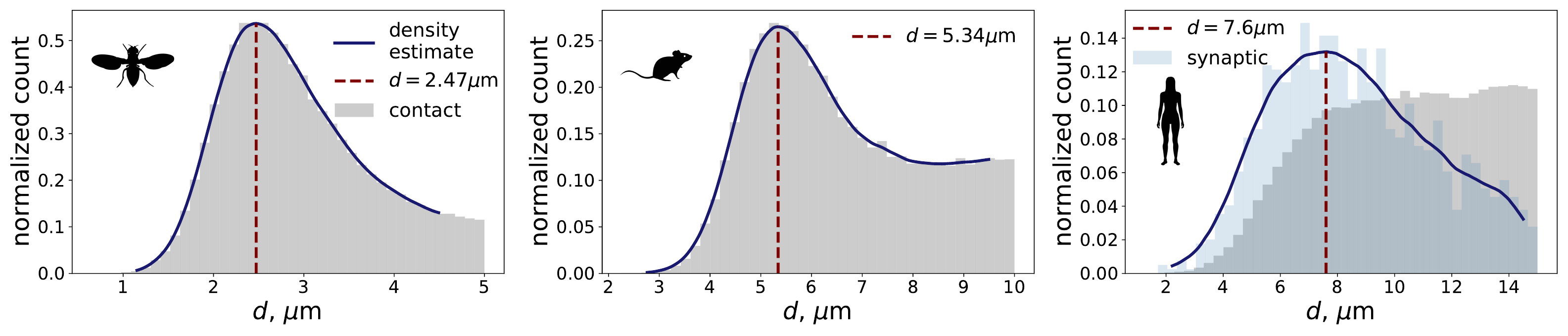}
    \caption{Soma size estimation. Grey: binned distances from contact to soma (only plotted for visualization purposes). Blue: kernel density estimation using linear kernel forms. Red: location of the peak of the density function that we use as a soma size. $d$ stands for the distance from contact to the center of the soma.}
    \label{fig: soma sizes}
\end{figure}

\iffalse \begin{figure}
    \centering
    \includegraphics[width=\linewidth]{Fig_SI_15.pdf}    \caption{\hl{Figure needs to be edited: can discuss.} Relationship between synaptic degree and neuron properties (number of vertices in the mesh and linear span).}
    \label{fig: deg corr}
\end{figure}
\fi

\iffalse\begin{figure}
    \centering
    \includegraphics[width=.7\linewidth]{figures_updated/mouse_angle_dependence.pdf}
    \caption{\textbf{Left}: distance dependence exponent $d_0$ for Euclidean distance in 3D vs Euclidean distances measured along the $x$, $y$, and $z$ axes. 
    The value along the $y$ axis is distinct from the more similar values along $x$ and $z$. This can be explained by the alignment of neurons in the cortical columns with their axons roughly in the $y$ direction. \textbf{Right}: the angle between the orientation of individual neurons and the $x$ axis in the $xy$ plane (upper sub-figure) and $z$ axis in the $yz$ plane (lower sub-figure). The peaks indicate most neurons are roughly aligned with the $y$ axis.}
    \label{fig: soma sizes}
\end{figure}\fi

\iffalse \begin{figure}
    \centering   \includegraphics[width=\linewidth]{figures_updated/mouse_orientation.pdf}
    \caption{
    Heterogeneity in distance dependence in the mouse network. Here, distance dependence is calculated along the vectors associated with the neuron orientation in space, as defined by the dominant PCA vector of their mesh vertices. Then, PCA was performed on the resulting dominant direction vectors. In this figure, we denote the principal component vectors obtained through this procedure as $x$, $y$, and $z$. The components are $x=(0.15,0.96,.23)$, $y=(-0.98,.15,0.04)$, and $z=(-0.,-0.23,0.97)$ in terms of the data coordinate system.
    The explained variance ratios are $0.63$, $0.23$, and $0.14$ for $x$, $y$, and $z$ directions respectively.
    \textbf{Upper left}: Euclidean distance dependence (grey) and the distance dependence in $x$, $y$, and $z$ directions (olive, dark blue, and dark yellow). The exponents are estimated to be $d=9$, $d_x=15$, $d_y=10$, and $d_z=8$ soma sizes.
    \textbf{Lower left}: distribution of angles between the dominant PCA vector and the 
    $x$ axis in the $xy$ plane ($\theta_{xy}$) and the $xz$ plane ($\theta_{xz}$). \textbf{Right}: distance dependence in models. Qualitatively, models that use the contact constraint $\textbf{c}$ reproduce the heterogeneity of the distance dependence in data.}    \label{fig: neuron orientation mouse}
\end{figure}
\fi

\begin{table}[ht]
\centering
\resizebox{\textwidth}{!}{%
\begin{tabular}[t]{c|c|c|c|c|c|c|c|c|c|c|c|c}
&\begin{tabular}{@{}c@{}}soma \\ size\end{tabular}
&\begin{tabular}{@{}c@{}}contact \\ thr.\end{tabular}
&\begin{tabular}{@{}c@{}}$\#$ \\ 
nodes\end{tabular}
&\begin{tabular}{@{}c@{}}network \\ type\end{tabular}
&\begin{tabular}{@{}c@{}}$\#$ connected \\ nodes\end{tabular}
&\begin{tabular}{@{}c@{}}$\#$nodes \\ in LCC\end{tabular}
&$\#$ edges
&\begin{tabular}{@{}c@{}}density, \\ $\%$\end{tabular}
&\begin{tabular}{@{}c@{}}max. \\ degree\end{tabular}
&\begin{tabular}{@{}c@{}}mean \\ degree\end{tabular}
&\begin{tabular}{@{}c@{}}median \\ degree\end{tabular}
&\begin{tabular}{@{}c@{}}wiring \\ len. ratio\end{tabular}\\
\hline
\multirow{2}{*}{\textbf{fly}}
&\multirow{2}{*}{2.47$\mu $m}
&\multirow{2}{*}{40nm}
& \multirow{2}{*}{16,804}
&synaptic&16,804&16,804&1,936,798&0.34&3,097&230.52&195&\multirow{2}{*}{0.24}\\
&&&&contact&16,804&16,804&7,747,751&1.37&7,335&922.13&806&\\
\hline

\multirow{2}{*}{\textbf{mouse}}
&\multirow{2}{*}{5.34$\mu $m}
&\multirow{2}{*}{41nm}
& \multirow{2}{*}{6,548}
&synaptic&6,261&6,224&111,636&0.13&881&34.10&23&\multirow{2}{*}{0.07}\\
&&&&contact&6,507&6,485&1,373,875&1.60&2,042&419.63&418\\
\hline

\multirow{2}{*}{\textbf{human}}
&\multirow{2}{*}{7.6$\mu $m}
&\multirow{2}{*}{46nm}
& \multirow{2}{*}{15,730}
&synaptic&13,579&13,352&74,934&0.0032&138&9.53&6&\multirow{2}{*}{0.03}\\
&&&&contact&15,129&15,001&2,049,143&0.41&1,221&260.54&188.5\\

\end{tabular}}
\caption{Properties of the synaptic and contact networks.}
\label{table: basic info}
\end{table}

\begin{table}[ht]
\centering
\resizebox{\textwidth}{!}{%
\begin{tabular}[t]{c|c|c|c|c|c|c|c|cccc|c|cccc}

organism
&$pc_1$
&$pc_2$
&$pc_3$
&EVR$_1$
&EVR$_2$
&EVR$_3$
&$d_{\text{tr}}$
&$d$
&$d_1$
&$d_2$
&$d_3$
&$d_{\text{tr}}^c$
&$d^c$
&$d_1^c$
&$d_2^c$
&$d_3^c$\\
\hline
\textbf{fly}&(-0.47,-0.84,-0.28)&(-0.75,0.22,0.62)&(-0.46,-0.50,-0.73)&0.39&
0.33&0.28&10&9&17&10&9&12&9&14&10&10\\
\textbf{mouse}&(0.15,0.96,0.23)&(-0.98,0.15,0.04)&(-0.00,-0.23,-0.97)&0.63&
0.23&0.14&10&9&15&10&8&12&10&17&10&8\\
\textbf{human}&(0.71,0.25,-0.08)&(-0.71,-0.7,0.08)&(-0.00,-0.11,0.99)&0.71&0.25&0.05&12&15&21&9&2&15&19&26&10&3\\
\end{tabular}}
\caption{Spatial orientation of neurons and distance dependence. Here, distance dependence is calculated along the vectors associated with the neuron orientation in space, as defined by the dominant PCA vector of their mesh vertices. Then, PCA was performed on the resulting dominant direction vectors, $pc_i$ correspond to the principal component vectors obtained using this procedure. EVR$_i$ are the corresponding explained variance ratios. $d$ is the characteristic distance estimated from $p(d')\propto e^{-d'/d}$ where $d'$ is the Euclidean distance between the neurons in 3D. $d_i$ are the characteristic distances for the 1D Euclidean distance along the principal component directions.}
\label{table: orientation}
\end{table}

\begin{table}[ht]
\centering
\resizebox{.9\textwidth}{!}{
%\begin{center}
\begin{tabular}{ c|c|c|c|c|c|c|c } 
 model & network 
& \begin{tabular}{@{}c@{}}AUC-ROC, \\ fly\end{tabular} 
& \begin{tabular}{@{}c@{}}AUC-ROC, \\ mouse\end{tabular} 
& \begin{tabular}{@{}c@{}}AUC-ROC, \\ human\end{tabular} 
& \begin{tabular}{@{}c@{}}AUC-PR, \\ fly\end{tabular} 
& \begin{tabular}{@{}c@{}}AUC-PR, \\ mouse\end{tabular} 
& \begin{tabular}{@{}c@{}}AUC-PR, \\ human\end{tabular} \\
\hline
 \textbf{d} & contact & 0.86&\textbf{0.95}&\textbf{0.96}&0.29&\textbf{0.62}&0.32\\ 
 \textbf{k} & contact & 0.70&0.73&0.79&0.15&0.17&0.07\\ 
 \textbf{k+L} & contact & \textbf{0.89}&\textbf{0.95}&\textbf{0.96}&\textbf{0.40}&\textbf{0.62}&\textbf{0.48}\\
 \hline
 \textbf{d} & synaptic & 0.85&0.94&0.97&0.10&0.07&0.03\\ 
 \textbf{k} & synaptic & 0.75&0.84&0.86&0.06&0.07&0.01\\ 
 \textbf{k+L} & synaptic & \textbf{0.90}&\textbf{0.96}&\textbf{0.99}&\textbf{0.18}&\textbf{0.25}&\textbf{0.09}\\
 \hline
  \textbf{d+c} & synaptic & 0.54&0.57&0.64&0.30&0.10&0.06\\ 
 \textbf{k+c} & synaptic & \textbf{0.68}&\textbf{0.75}&\textbf{0.73}&\textbf{0.42}&\textbf{0.24}&\textbf{0.12}\\
\end{tabular}}
\caption{List of AUC-ROC and AUC-PR values for predicting the network edges from models. Rows 1-3: predicting contactome edges from connectome models, see \cref{fig: ROC and PR} for the ROC and precision-recall curves. Rows 4-6: predicting connectome edges from connectome models, see \cref{fig: ROC and PR syn}. Rows 7,8: predicting connectome edges from connectome models restricted by the contactome, see \cref{fig: ROC and PR syn cont}. The values corresponding to the model with the best prediction within each category are shown in bold font for each organism.}
\label{table: AUC}
%\end{center}}
\end{table}

\begin{table}[ht]
\centering
\resizebox{.42\textwidth}{!}{
%\begin{center}
\begin{tabular}{ c|c|c|c } 
 model 
& fly
& mouse
& human \\
\hline
 \textbf{d} & -8,350,235&-498,369&-429,298\\ 
 \textbf{k} & -9,286,707&-579,990&-548,372\\ 
 \textbf{k+L} & \textbf{-7,441,092} &\textbf{-418,911}&\textbf{-364,679}\\
 \hline
 \textbf{c} & -4,356,676 & -387,194& -321,470\\ 
 \textbf{d+c} &  -4,323,885 & -384,247&-312,048\\ 
 \textbf{k+c} & \textbf{-4,055,652} & \textbf{-342,900} &\textbf{-292,252} \\
\end{tabular}}
\caption{Log-likelihood values of the empirical connectome for different models. The values corresponding to the model with the highest log-likelihood within each category are shown in bold font for each organism.}
\label{table: likelihood}
%\end{center}}
\end{table}

\begin{table}[ht]
\centering
\resizebox{\textwidth}{!}{%
\begin{tabular}[t]{c|c|c|c|c|c|c|c|c|c|c|c}

organism
&\begin{tabular}{@{}c@{}}$\#$nodes \\ in LCC\end{tabular}
&diam.
&\begin{tabular}{@{}c@{}}average \\ sh. path\end{tabular}
&\begin{tabular}{@{}c@{}}global \\ efficiency\end{tabular}
&\begin{tabular}{@{}c@{}}local \\ efficiency\end{tabular}
&\begin{tabular}{@{}c@{}}transi- \\ tivity\end{tabular}
&\begin{tabular}{@{}c@{}}clust. \\ coeff.\end{tabular}
&$\#$ triangles
&\begin{tabular}{@{}c@{}}$\#$ squares, \\ no diagonal\end{tabular}
&\begin{tabular}{@{}c@{}}$\#$ squares, \\ one diagonal\end{tabular}
&\begin{tabular}{@{}c@{}}$\#$ squares, \\ two diagonals\end{tabular}\\
\hline
\textbf{fly}&16,804&5&2.53&0.42&
0.64&0.21&0.30&52,367,260&2,616,817,833&5,405,355,056&781,627,819\\
\textbf{mouse}&6,224&12&3.32&0.30&
0.26&0.10&0.11&366,256&8,804,181&14,863,110&863,837\\
\textbf{human}&13,352&21&5.60&0.14&0.06&0.05&0.04&24,242&253,578&71,290&2,012
\end{tabular}}
\caption{Properties of the synaptic and contact networks used in \cref{fig: fold changes,fig: fold changes extra}. For definitions of network metrics used in this table, see \cref{subsec: network measures}.}
\label{table: less basic info}
\end{table}

\begin{table}[ht]
\centering
\resizebox{.7\textwidth}{!}{
%\begin{center}
\begin{tabular}{ c|c|c } 
 name & soft constraints & contact constraint\\ 
 \hline
 \textbf{d} & binned edge probability as a function of distance&no\\ 
 \textbf{k} & degree sequence&no \\ 
 \textbf{k+L} & degree sequence and total edge length&no \\
 \hline
  \textbf{c} & -&yes \\
  \textbf{d+c} &binned edge probability as a function of distance &yes \\
  \textbf{k+c} &degree sequence &yes \\
\end{tabular}}
\caption{List of maximum entropy models we use throughout the paper. Note that we only consider the maximum entropy models that, on average, preserve the total number of edges in the network.}
\label{table: randomizations}
%\end{center}}
\end{table}

\end{document}